\documentclass[usenatbib,useAMS]{mn2e}
\bibpunct{(}{)}{;}{a}{}{,}
\usepackage[dvips]{graphicx}
\usepackage{times}
\usepackage{color}
\usepackage{appendix}
\usepackage{tabularx}
\usepackage{array}

\renewcommand{\theequation}{\thesection.\arabic{equation}}

\newcommand{\mlr}{\hbox{$M_\ast/ L_r$}}
\newcommand{\mlrsun}{\hbox{$M_\odot/ L_{r,\odot}$}}
\newcommand{\fsfh}{\hbox{$f_\mathrm{SFH}$}}
\newcommand{\sfr}{\hbox{$\psi$}} 
\newcommand{\ssfr}{\hbox{$\psi_{\mathrm S}$}}
\newcommand{\zgas}{\hbox{$Z$}}
\newcommand{\hii}{\mbox{H\,{\sc ii}}}
\newcommand{\tauv}{\hbox{$\hat{\tau}_{V}$}}
\newcommand{\tauvbc}{\hbox{$\hat{\tau}_{V}^{\mathrm{BC}}$}}
\newcommand{\tauvism}{\hbox{$\hat{\tau}_{V}^{\mathrm{ISM}}$}}
\newcommand{\taul}{\hbox{$\hat{\tau}_\lambda$}}
\newcommand{\taulbc}{\hbox{$\hat{\tau}_\lambda^{\mathrm{BC}}$}}
\newcommand{\taulism}{\hbox{$\hat{\tau}_\lambda^{\mathrm{ISM}}$}}
\newcommand{\ha}{\hbox{H$\alpha$}}
\newcommand{\hb}{\hbox{H$\beta$}}
\newcommand{\mstar}{\hbox{$M_\star$}}

\newcommand{\xid}{\hbox{$\xi_{\mathrm d}$}}

\newcommand{\msun}{\hbox{$M_\odot$}}
\newcommand{\zsun}{\hbox{$Z_\odot$}}
\newcommand{\oii}{\hbox{[O\,{\sc ii}]}}
\newcommand{\oiii}{\hbox{[O\,{\sc iii}]}}
\newcommand{\nii}{\hbox{[N\,{\sc ii}]}}
\newcommand{\sii}{\hbox{[S\,{\sc ii}]}}
\newcommand{\sn}{\hbox{\textrm{S/N}}}
\newcommand{\snmed}{\hbox{$\overline{\textrm{S/N}}$}}
\newcommand{\oh}{\hbox{\textrm{O/H}}}
\newcommand{\aboh}{\hbox{$12+\log\textrm{(O/H)}$}}
\newcommand{\isig}{\hbox{$I_\sigma$}}
\newcommand{\zobs}{\hbox{$z_\mathrm{obs}$}}

\newcommand{\hdg}{\hbox{H$\delta_A$+H$\gamma_A$}}
\newcommand{\mgfep}{\hbox{$\textrm{[MgFe]}^\prime$}}
\newcommand{\mgtfe}{\hbox{$\textrm{[Mg}_2\textrm{Fe]}$}}
\newcolumntype{x}[1]{%
>{\centering\hspace{0pt}}p{#1}}%
\newcommand{\tn}{\tabularnewline}

\title[Constraining galaxy physical parameters]
{Relative merits of different types of rest-frame optical observations to constrain galaxy physical parameters}

\author[C. Pacifici et al.]
{Camilla~Pacifici,$^{1,2}$\thanks{E-mail: pacifici@iap.fr} St\'ephane~Charlot,$^{1,2}$ J\'er\'emy~Blaizot$^{3,4,5}$ and Jarle~Brinchmann$^{6}$
\\
$^{1}$UPMC Universit\'e Paris 6, UMR7095, Institut d'Astrophysique de Paris, F-75014, Paris, France\\
$^{2}$CNRS, UMR7095, Institut d'Astrophysique de Paris, F-75014, Paris, France\\
$^{3}$Universit\'e de Lyon, Lyon F-69003; Universit\'e de Lyon 1, Observatoire de Lyon, Saint-Genis Laval F-69230, France\\
$^{4}$CNRS, UMR5574, Centre de Recherche Astrophysique de Lyon, Lyon F-69007, France\\
$^{5}$Ecole Normale Sup\'erieure de Lyon, Lyon F-69007, France\\
$^{6}$Leiden Observatory, Leiden University, PO Box 9513, 2300 RA Leiden, The Netherlands
}

\begin{document}

\maketitle

\begin{abstract}
We present a new approach to constrain galaxy physical parameters from the combined interpretation of stellar and nebular emission in wide ranges of observations. This approach relies on the Bayesian analysis of any type of galaxy spectral energy distribution using a comprehensive library of synthetic spectra assembled using state-of-the-art models of star formation and chemical enrichment histories, stellar population synthesis, nebular emission and attenuation by dust. We focus on the constraints set by 5-band $ugriz$ photometry and low- and medium-resolution spectroscopy at rest wavelengths $\lambda$=3600--7400\,\AA\ on a few physical parameters of galaxies: the observer-frame absolute $r$-band stellar mass-to-light ratio, \mlr; the fraction of a current galaxy stellar mass formed during the last 2.5\,Gyr, \fsfh;  the specific star formation rate, \ssfr; the gas-phase oxygen abundance, \aboh; the total effective $V$-band absorption optical depth of the dust, \tauv; and the fraction of this arising from dust in the ambient interstellar medium, $\mu$. Since these parameters cannot be known a priori for any galaxy sample, we assess the accuracy to which they can be retrieved from observations by simulating `pseudo-observations' using models with known parameters. Assuming that these models are good approximations of true galaxies, we find that the combined analysis of stellar and nebular emission in low-resolution (50\,\AA\ FWHM) galaxy spectra provides valuable constraints on all physical parameters. The typical uncertainties in high-quality spectra are about 0.13\,dex for \mlr, 0.23 for \fsfh, 0.24\,dex for \ssfr, 0.28 for \aboh, 0.64 for \tauv\ and 0.16 for $\mu$. The uncertainties in \aboh\ and \tauv\ tighten by about 20 percent for galaxies with detectable emission lines and by another 45 percent when the spectral resolution is increased to 5\,\AA\  FWHM. At this spectral resolution, the analysis of the combined stellar and nebular emission in the high-quality spectra of 12,660 SDSS star-forming galaxies using our approach yields likelihood distributions of \mstar, \aboh, \tauv\ and \ssfr\ similar to those obtained in previous separate analyses of the stellar and nebular emission at the original (twice higher) SDSS spectral resolution. Meanwhile, rest-frame $ugriz$ photometry provides competitive constraints on \mlr. We show that the constraints derived on galaxy physical parameters from these different types of observations depend sensitively on signal-to-noise ratio. Our approach can be extended to the analysis of any type of observation across the wavelength range covered by spectral evolution models.
\end{abstract}


\section{Introduction}
\label{sec:intro}

The light emitted by stars and the interstellar medium in galaxies contains important clues about the physical processes that drive galaxy evolution. Different strategies have been adopted to survey the spectral energy distributions of large samples of galaxies to gather these clues. High-quality photometry from surveys such as the Sloan Digital Sky Survey (SDSS, \citealt{york2000}), the Canada-France-Hawaii Telescope Legacy Survey (CFHTLS,  \citealt{goranova2009}) and the Cosmic Evolution Survey (COSMOS, \citealt{scoville2007}) has brought important information about the rest-frame optical emission from millions of galaxies out to redshifts of a few. This is supplemented by medium-resolution spectroscopy of over a million local galaxies from the SDSS and the Two-degree Field Galaxy Redshift Survey (2dFGRS, \citealt{colless2001}), and low- and medium-resolution spectroscopy of tens of thousands of galaxies at higher redshift from surveys such as the VIMOS VLT Deep Survey (VVDS, \citealt{lefevre2005}) and the Deep Extragalactic Evolutionary Probe (DEEP2, \citealt{davis2003}). The interpretation of the high-quality data gathered by these surveys has triggered the development of sophisticated models of galaxy spectral evolution (e.g., \citealt{bruzual2003, gonzalez2005, maraston2005, coelho2007,walcher2009, vazdekis2010,thomas2011}; see also \citealt{charlot2001, kewley2002}). These models can help us  constrain the star formation and chemical enrichment histories of galaxies through the analysis of different types of photometric and spectroscopic observations. So far, however, the relative merits of different observational approaches to constrain basic physical properties of galaxies have not been quantified in detail.

Recent studies have brought important insight into the efficiency of different types of observations to constrain physical quantities such as star formation history, metallicity and dust content from observations of the rest-frame optical emission of galaxies. Multiband photometry has proved valuable to constrain galaxy redshifts and basic quantities such as stellar mass-to-light ratio (provided a stellar initial mass function; hereafter IMF) and, when combined with rest-frame ultraviolet and near-infrared constraints, star formation rate and attenuation by dust (e.g., \citealt{bell2001, salim2005, salim2007, ilbert2006, schaerer2010}; but see \citealt{wuyts2009, conroy2010}). Medium-resolution spectroscopy allows separate analyses of nebular-emission and stellar-absorption features, and hence, more accurate constraints on the star formation history, stellar and interstellar metallicities, attenuation by dust and nuclear activity (e.g., \citealt{kauffmann2003a, brinchmann2004, heckman2004, tremonti2004, cid2005,gallazzi2005, panter2007, wild2007}). At low spectral resolution, the difficulty of distinguishing between the signatures of nebular-emission and stellar-absorption features, especially around H-Balmer lines, renders analyses more complicated \citep[e.g.,][]{lamareille2006}. This difficulty could be overcome with the ability to interpret simultaneously the stellar and nebular emission from galaxies. In this case, the relative merits of multiband photometry and low- and medium-resolution spectroscopy to constrain galaxy physical parameters would need to be reexamined.

In this paper, we develop a new approach to interpret the combined stellar and nebular emission from galaxies in wide ranges of multi-wavelength observations. To achieve this, we appeal to a set of state-of-the-art models of galaxy star formation and chemical enrichment histories, stellar population synthesis, nebular emission and attenuation by dust. We use these models to generate a comprehensive library of several million galaxy spectral energy distributions. This library can be used to retrieve probability density functions of physical parameters from the Bayesian analysis of any observed spectral energy distribution. We focus here on the constraints set by 5-band $ugriz$ photometry and low- and medium-resolution spectroscopy at rest wavelengths $\lambda$=3600--7400\,\AA\ on a few basic physical parameters of galaxies: the observer-frame absolute $r$-band stellar mass-to-light ratio (\mlr); the fraction of a current galaxy stellar mass formed during the last 2.5\,Gyr (\fsfh);  the specific star formation rate (\ssfr); the gas-phase oxygen abundance [\aboh]; the total effective $V$-band absorption optical depth of the dust (\tauv); and the fraction of this arising from dust in the ambient interstellar medium (hereafter ISM; $\mu$). These parameters cannot be known a priori for any galaxy sample. To assess the accuracy to which they can be retrieved from different types of observations, we therefore simulate `pseudo-observations' by convolving the spectral energy distributions of models with known parameters with appropriate instrument responses and then applying artificial noise to mimic true observations. This approach allows us to make accurate predictions for the optimistic case in which the models we rely on are good approximations of true galaxies. A most notable outcome of our study is that the combined analysis of stellar and nebular emission in low-resolution (50\,\AA\  FWHM), high-quality (median signal-to-noise ratio per pixel $\snmed=20$) galaxy spectra provides valuable constraints on all the physical parameters mentioned above. We also explore how the constraints from different types of photometric and spectroscopic observations depend on signal-to-noise ratio and the inclusion of nebular emission in the analysis. Our main results are summarized in Table~\ref{tab:summary} (Section~\ref{sec:conclusion}).

We describe our approach to constrain galaxy physical parameters from the combined analysis of stellar and nebular emission in Section \ref{sec:models} below. In Section \ref{sec:estim}, we investigate the accuracy to which the parameters  \mlr, \fsfh, \ssfr, \aboh, \tauv\ and $\mu$ can be constrained from different types of photometric and spectroscopic observations. Then, in Section \ref{sec:sample}, we compare the constraints derived from the analysis of the medium-resolution spectra of 12,660 SDSS star-forming galaxies using our approach with those obtained in previous separate analyses of the stellar and nebular emission. We also investigate the influence of the prior distributions of physical parameters on the retrieved probability density functions. Our conclusions are summarized in Section \ref{sec:conclusion}. Throughout this paper, we adopt the standard cosmology $\Omega_{\mathrm{M}}=0.3$, $\Omega_{\Lambda}=0.7$ and $h=0.7$.


\section{Modeling approach}
\label{sec:models}
In this section, we describe our approach to assess the retrievability of galaxy physical parameters from the combined analysis of stellar and nebular emission in wide ranges of multi-wavelength observations. We first build a comprehensive library of star formation and chemical enrichment histories of galaxies (Section~\ref{sec:millennium}). We appeal to state-of-the-art models of spectral evolution to compute the emission from stars and gas and the attenuation by dust in these galaxies (Section~\ref{sec:spectra}). This allows us to assemble a large library of galaxy spectral energy distributions (Section~\ref{sec:cats}), which we can use to retrieve likelihood distributions of physical parameters from the Bayesian analysis of any type of observed spectral energy distribution (Section~\ref{sec:stat}).


\subsection{Library of star formation and chemical enrichment histories}
\label{sec:millennium}

\begin{figure}
\begin{center}
\includegraphics[width=0.45\textwidth]{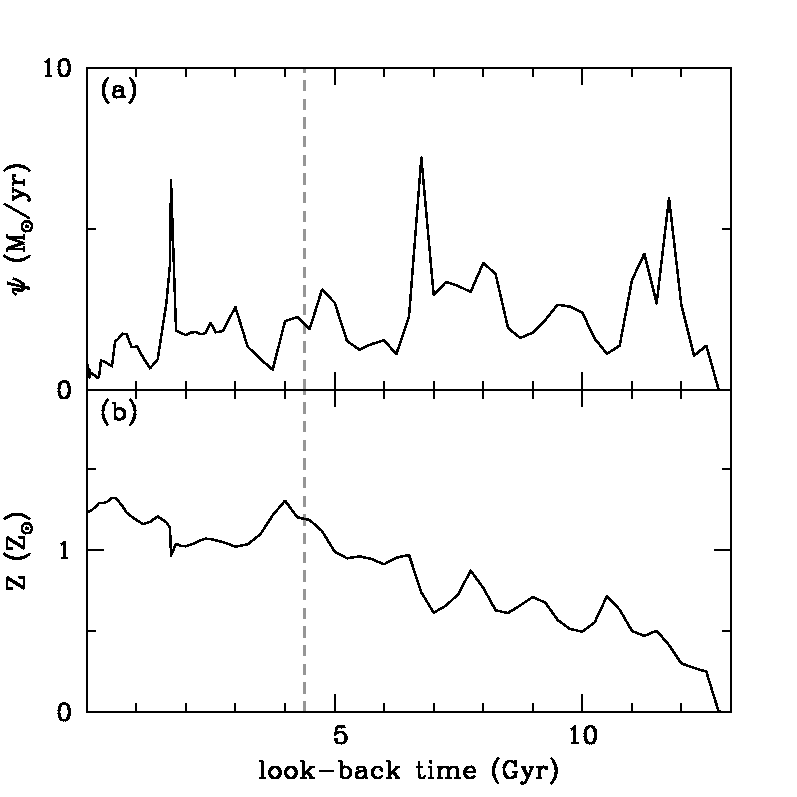}
\caption{Example of galaxy star formation and chemical enrichment histories inferred from the semi-analytic post-treatment of the Millennium cosmological simulation: ({\it a}) star formation rate, \sfr, and ({\it b}) interstellar metallicity, \zgas, plotted as a function of look-back time, for a galaxy with a present-day stellar mass of $1.5\times10^{10}\msun$. The vertical dashed line marks the evolutionary stage at which the galaxy is looked at in Fig.~\ref{fig:methods} (corresponding to a galaxy age of 8.4\,Gyr).}
\label{fig:sfhmfh}
\end{center}
\end{figure}

We build a comprehensive library of galaxy star formation and chemical enrichment histories by performing a semi-analytic post-treatment of the Millennium cosmological simulation of \citet{springel2005}. This large-scale cosmological simulation follows the growth, interaction and merging history of dark matter haloes from redshift $z=127$ to the present time.\footnote{In the Millennium simulation, dark matter is represented as a set of about $10^{10}$ point particles, which interact gravitationally in a cubic region of side $500h^{-1}\,$Mpc. Individual particles have a mass of $8.6 \times 10^8 h^{-1}$\msun, and the detection threshold for a dark matter halo is 20 particles, i.e.  about $1.7 \times 10^{10} h^{-1}$\msun\ \citep[see][for detail]{springel2005}.} We adopt the simple semi-analytic recipes of \citet[][see also \citealt{croton2006}]{delucia2007} to follow star formation and the associated metal production by gas falling into these dark matter haloes. We compute in this way the star formation and chemical enrichment histories of 500,000 galaxies with present-day stellar masses comprised between $\mstar\approx 5\times10^9 h^{-1}\msun$ (corresponding roughly to the mass resolution of the simulation; \citealt{blaizot2004}) and $5\times10^{11} h^{-1}\msun$. Figure~\ref{fig:sfhmfh} shows an example of star formation and chemical enrichment histories of a galaxy with present-day stellar mass $1.5\times10^{10}\msun$. The spikes in both curves reflect the various episodes of interactions and mergers that shaped the evolution of the galaxy. As shown by \citet{wild2008}, star formation histories computed in this way account well for the observed properties of galaxies at different redshifts in the VVDS ($0.5\la z\la1$) and SDSS ($0.05\la z\la0.1$) samples. 

About 35 percent of the 500,000 galaxies generated from the semi-analytic post-treatment of the Millennium simulation do not form stars at $z=0$. These contain a negligible amount of interstellar gas and correspond to the population of present-day early-type galaxies. The remaining galaxies span a relatively narrow range of star formation properties. Their specific star formation rates peak around $\ssfr\approx0.1\,\mathrm{Gyr}^{-1}$, with a dispersion of only a factor of 2. Furthermore, the interstellar metallicity \zgas\ in these galaxies correlates significantly with \ssfr, in the sense that the most metal-poor galaxies form stars most actively. The galaxies also have roughly similar recent histories of star formation, as quantified by the star formation history parameter \fsfh, which we define as the fraction of the current galaxy stellar mass formed during the last 2.5\,Gyr (this time lag corresponds to the main-sequence lifetime of A-F stars with strong Balmer absorption features). This parameter has a narrow distribution  centred on $\fsfh\approx0.15$, with a dispersion of only 0.1. The distributions of \ssfr\ and \fsfh\ are compatible with roughly constant (overall) star formation over the past 2.5\,Gyr in galaxies more massive than $5\times10^9 h^{-1}\msun$, given that 40 percent of the stellar mass formed during this period is returned to the ISM in the form of winds and supernovae \citep[e.g.][]{bruzual2003}. The above distributions of \ssfr, \zgas\ and \fsfh\ are consistent with the observed properties of nearby, star-forming SDSS galaxies \citep[e.g.][]{kauffmann2003a,brinchmann2004,tremonti2004}. 

The relatively limited ranges of star formation histories and interstellar metallicities described above, while appropriate to interpret observations of nearby SDSS galaxies more massive than $5\times10^9 h^{-1}\msun$, could bias statistical estimates of \ssfr, \zgas\  and \fsfh\ in galaxies drawn from other samples or in different mass ranges (e.g. \citealt{weinmann2006}; see also Section~\ref{sec:prior} below). To avoid such biases, we broaden the original prior distributions of these parameters predicted by the semi-analytic recipes of \citet{delucia2007} in 2 ways. Firstly, we widen the range in \fsfh\ by randomly drawing the evolutionary stage (i.e. galaxy age) at which a galaxy is looked at in the library. In practice, we draw this stage uniformly in redshift between $z=0$ and 1.7, i.e. the upper limit beyond which \fsfh\ reaches unity for a substantial fraction of the galaxies (Fig.~\ref{fig:dist}b below; this resampling makes the proportion of early-type galaxies drop from 35 to about 10 percent in the library). Secondly, for galaxies with ongoing star formation at the selected stage, we broaden the distributions of the current star formation rate and the current interstellar metallicity, as follows (we define here as `current' the average of a quantity over a period of 10\,Myr before a galaxy is looked at). We redistribute the specific star formation rate over the interval $-2<\log(\ssfr/\mathrm{Gyr}^{-1})< 1$ using the probability density function $p[\log(\ssfr/\mathrm{Gyr}^{-1})]=0.05[1.5^5-\left|\log(\ssfr/\mathrm{Gyr}^{-1})+0.5\right|^5]$, which is approximately uniform over the logarithmic range from  $\ssfr=0.03$ to $3\,\mathrm{Gyr}^{-1}$ and drops to zero at $\ssfr=0.01$ and $10\,\mathrm{Gyr}^{-1}$ (Fig.~\ref{fig:dist}c). We redistribute the current interstellar metallicity \zgas\ logarithmically between 0.1 and 3.5 times solar ($\zsun=0.017$). 

The random resampling of \ssfr\ and \zgas\ for star-forming galaxies in the library is justified in part by the stochastic nature of star formation and chemical evolution in hierarchical scenarios of galaxy formation. Also, since the changes impact only the last 10\,Myr of evolution, they have a negligible influence on the global star formation and chemical enrichment histories derived from the semi-analytic post-treatment of the Millennium simulation. In return, the use of wider and uncorrelated prior distributions of \ssfr\ and \zgas, together with the broader sampling in galaxy age (and \fsfh), greatly enhances the usefulness of the model library to explore and interpret the spectral properties of observed galaxy populations.


\subsection{Galaxy spectral modeling}
\label{sec:spectra}
We now describe the way in which we compute the emission from stars and gas and the attenuation by dust in the model galaxies generated in the previous section. The luminosity per unit wavelength $L_\lambda (t)$ emerging at time $t$ from a galaxy can be expressed as \citep[see, e.g.,][]{charlot2001}
\begin{equation}
\label{eq:lum}
L_{\lambda} (t) = \int_{0}^{t} dt'\, \sfr(t-t') \,  S_{\lambda}[t',\zgas(t-t')] \, T_{\lambda}(t,t') \,,
\end{equation}
where $\psi(t-t')$ is the star formation rate at time $t-t'$, $S_{\lambda}[t',\zgas(t-t')]$ is the luminosity emitted per unit wavelength and per unit mass by a stellar generation of age $t'$ and metallicity $\zgas(t-t')$, and  $T_{\lambda}(t,t')$ is the transmission function of the interstellar medium, defined as the fraction of the radiation produced at wavelength $\lambda$ at time $t$ by a generation of stars of age $t'$ that is transferred by the ISM. Following \citet{charlot2001}, we write $T_{\lambda}(t,t')$ as the product of the transmission functions of the ionized gas, $T_{\lambda}^+(t,t')$, and the neutral ISM, $T_{\lambda}^0(t,t')$,
\begin{equation}
\label{eq:tfunc}
T_{\lambda}(t,t')=T_{\lambda}^+(t,t')T_{\lambda}^0(t,t')\,.
\end{equation}
In the next paragraphs, we describe the prescriptions we adopt for the functions $S_{\lambda}$, $T_{\lambda}^+$ and $T_{\lambda}^0$.

\subsubsection{Stellar emission}
\label{sec:stars}
We compute the luminosity $S_{\lambda}(t,\zgas)$ emitted per unit wavelength and per unit mass by a stellar generation of age $t$ and metallicity $\zgas$ using the latest version of \cite{bruzual2003} stellar population synthesis code (Charlot \& Bruzual, in preparation). This code predicts the spectral evolution of stellar populations at wavelengths from 91\,{\AA} to 160\,$\mu$m and at ages between $1\times10^5$ and $2\times10^{10}\,$yr, for different metallicities (from about 0.005\zsun\ to 4\zsun), initial mass functions and star formation histories. We use the most recent version of the code, which incorporates a new library of observed stellar spectra \citep{sanchez2006} and new prescriptions for the evolution of stars less massive than 20\,\msun\  \citep{bertelli2008,bertelli2009} and for the thermally pulsating asymptotic giant branch (TP-AGB) evolution of low- and intermediate-mass stars \citep{marigo2008}. The main effect of the revised TP-AGB prescription is to improve the predicted near-infrared colours of intermediate-age stellar populations \citep[e.g.,][]{gonzalez2010}. In all applications throughout this paper, we adopt the Galactic-disc IMF of \citet{chabrier2003}.


\subsubsection{Nebular emission}
\label{sec:neb}
A main feature of our study is that we account for the contribution by ionized gas to the emission from galaxies. We follow the prescription of \cite{charlot2001} to compute the transmission function $T_{\lambda}^+(t,t')$ of the gas photoionized at time $t$ by stars of age $t'$ in a galaxy. This consists in adopting effective (i.e. galaxy-wide) parameters to describe the ensemble of \hii\ regions and the diffuse gas ionized by young stars throughout the galaxy. The main adjustable parameters are the interstellar metallicity, \zgas, the zero-age ionization parameter, $U_0$ (which characterizes the volume-averaged ratio of ionizing-photon to gas densities at age $t'=0$), and the dust-to-metal (mass) ratio, \xid\ (which characterizes the depletion of metals on to dust grains) of the photoionized gas.\footnote{These effective parameters were denoted by $\tilde{Z}$, $\langle\tilde{U}_0\rangle$ and  $\tilde{\xi}_{\mathrm d}$ in \citet{charlot2001}.} 

Following \citet{charlot2001}, we neglect the contribution by stars older than 10\,Myr to nebular emission, i.e., we write
\begin{equation}
T^+_{\lambda}(t,t')=1, \hskip1cm\mathrm{for}\hskip0.5cm t'>10\,\mathrm{Myr}\,.
\end{equation}
We use the standard photoionization code {\sc cloudy} \citep{ferland1996} to compute the transmission function at earlier ages, assuming that galaxies are ionization bounded. For $t'<10^7\mathrm{yr}$, therefore, $T^+_{\lambda}(t,t')$  is close to zero at wavelengths blueward of the H-Lyman limit and greater than unity at wavelengths corresponding to emission lines (we record the luminosities of the 109 most prominent emission lines at all wavelengths; we also include the recombination continuum radiation). When computing $T^+_{\lambda}(t,t')$ in equations~(\ref{eq:lum})--(\ref{eq:tfunc}), we take the metallicity of the photoionized gas to be the current metallicity $\zgas(t)$ of the galaxy. We randomly draw the dust-to-metal ratio from a uniform prior distribution between $\xid=0.1$ and $0.5$, and the zero-age ionization parameter from a logarithmic prior distribution between $\log\,U_0=-3.5$ and $-1.5$. We fix all other parameters of the photoionization code at the standard values favored by \citet[][see their paper for detail]{charlot2001}.

\begin{figure}
\begin{center}
\includegraphics[width=0.50\textwidth]{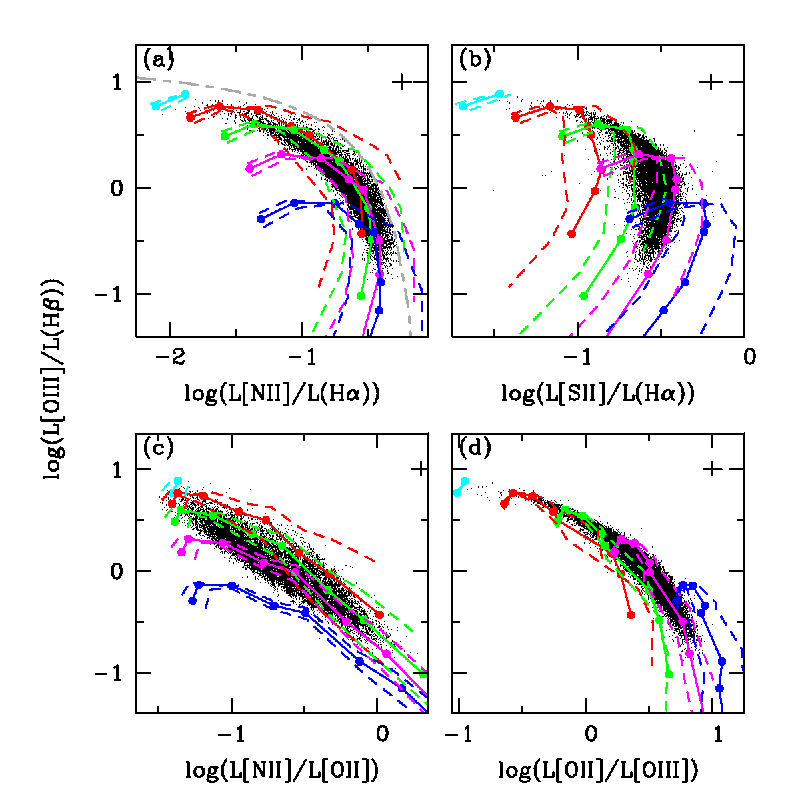}
\caption{Emission-line luminosities of galaxies computed using the models described in Section~\ref{sec:neb} (lines), compared with high-quality observations of a sample of 28,075 star-forming galaxies from the SDSS DR7 (black dots). The models assume constant star formation over the past 10\,Myr. The data are corrected for attenuation by dust as described in \citet{brinchmann2004}. ({\it a}) $L(\oiii\lambda5007)/L(\hb)$ versus $L(\nii\lambda6584)/L(\ha)$. ({\it b}) $L(\oiii\lambda5007)/L(\hb)$ versus $L(\sii\lambda\lambda6716,6731)/L(\ha)$. ({\it c}) $L(\oiii\lambda5007)/L(\hb)$ versus $L(\nii\lambda6584)/L(\oii\lambda3727)$. ({\it d}) $L(\oiii\lambda5007)/L(\hb)$ versus $L(\oii\lambda3727)/L(\oiii\lambda5007)$. In each panel, lines of different colours refer to models with different zero-age ionization parameter $\log U_0$ (cyan: $-1.5$;  red: $-2.0$;  green: $-2.5$; magenta: $-3.0$; blue: $-3.5$). At fixed $\log U_0$, the lower dashed line corresponds to models with dust-to-metal ratio $\xid=0.1$, the solid line to models with  $\xid=0.3$ and the upper dashed line to models with $\xid=0.5$. Along each line, dots mark the positions of models with different gas metallicity \zgas\ (0.10, 0.20, 0.50, 0.75, 1.00, 1.50, 2.00 and 3.00 times \zsun, from left to right; only low-metallicity models are considered for $\log U_0=-1.5$). In ({\it a}), the gray long-short-dashed line shows the \citet{kauffmann2003b} criterion to separate star-forming galaxies from AGNs.
}
\label{fig:lineratios}
\end{center}
\end{figure}

In Fig.~\ref{fig:lineratios}, we compare the luminosities of prominent optical emission lines obtained in this way for galaxies with different \zgas, $U_0$ and \xid, with high-quality observations of a sample of 28,075 star-forming galaxies from the SDSS Data Release 7 \citep[DR7;][]{abazajian2009}. The different panels show the relations between different luminosity ratios constructed with the $\oii\lambda\lambda3726,3729$ (hereafter $\oii\lambda3727$), \hb, $\oiii\lambda5007$, \ha, $\nii\lambda6584$ and $\sii\lambda\lambda6716,6731$ lines. We selected the star-forming galaxies of Fig.~\ref{fig:lineratios} from the SDSS DR7 by requiring high signal-to-noise ratio ($\sn>10$) in all lines and excluding galaxies in which the emission could be contaminated by an active galactic nucleus (AGN; we applied the conservative criterion of \citealt{kauffmann2003b} to reject AGNs from the standard \citealt{baldwin1981} line-diagnostic diagram reproduced in Fig.~\ref{fig:lineratios}a). The excellent agreement between models and SDSS observations in Fig.~\ref{fig:lineratios} is similar to that obtained by \citet{brinchmann2004} using the original \citet{charlot2001} models. This allows the derivation of  independent constraints on \zgas, $U_0$ and \xid\ from the different line luminosity ratios. For simplicity, in what follows, we adopt a fixed intrinsic line velocity dispersion of 100\,km\,s$^{-1}$ (typical of SDSS star-forming galaxies; e.g. \citealt{brinchmann2004}) when computing the contribution by ionized gas to the emission from galaxies.

\subsubsection{Attenuation by dust}
\label{sec:dust}
The  attenuation by dust of the emission from a galaxy depends on several factors, including the amount and structure of the interstellar gas, the metallicity and physical conditions of this gas and the orientation of the galaxy. Two galaxies with identical star formation and chemical enrichment histories but different sizes (and hence different total gas masses), ISM structures and orientations can have widely different dust attenuation optical depths, even though their dust-free spectral energy distributions are the same (modulo a mass scaling factor). We assume here for simplicity that the dust attenuation optical depth of a galaxy does not depend on other global galaxy parameters, such as the age $t$, the specific star formation rate \ssfr\ and the gas metallicity \zgas. We therefore write the transmission function of the neutral ISM in equations~(\ref{eq:lum})--(\ref{eq:tfunc}) as simply a function of stellar age $t'$,
\begin{equation}
T_{\lambda}^0(t,t')=T_{\lambda}^0(t')\,.
\label{eq:trans0}
\end{equation}

In the ISM, stars are born in giant molecular clouds, which dissipate on a time-scale of the order of 10\,Myr \citep[e.g.,][]{blitz1980,kawamura2009}. Thus the emission from stars younger than this time-scale and from the photoionized gas is more attenuated than that from older stars. We account for this effect using the simple dust model of \citet{charlot2000}, who express the transmission function of the neutral gas as\footnote{The function denoted here by $T_{\lambda}^0 (t')$ is equivalent to the function $T_{\lambda}(t')$ defined by equation (6) of \citet{charlot2000}, for $f=0.0$ in their notation.}
\begin{equation}
T_{\lambda}^0 (t')=\exp\left[-\taul (t')\right]\,.
\end{equation}
Here $\taul(t')$ is the `effective absorption' (or attenuation) optical depth of the dust  seen by stars of age $t'$. This represents the average absorption of photons emitted in all directions by stars in all locations within the galaxy, including the effects of scattering on the path lengths of the photons before they are absorbed. Following \citet{charlot2000}, we write
\begin{equation}
\label{eq:taul}
\taul(t')=\left\{ \begin{array}{l l}
\taulbc + \taulism & \hspace{3mm} \mathrm{for} \hspace{3mm} t' \leqslant 10\,\mathrm{Myr}\,,\\
\taulism & \hspace{3mm} \mathrm{for} \hspace{3mm}  t'>10\,\mathrm{Myr}\,, \end{array}\right.
\end{equation}
where \taulbc\ and \taulism\ are the effective absorption optical depths of the stellar birth clouds and the ambient (i.e. diffuse) ISM, respectively. We note that, since the intrinsic spectral energy distributions of young and old stars evolve in time, equation~(\ref{eq:taul}) implies that the global attenuation curve of a galaxy containing several generations of stars is complicated function of galaxy age, even though $\taul$ is a simple function of $t'$ (see, e.g., fig.~5 of \citealt{charlot2000}).

We must also specify the dependence of \taulbc\ and \taulism\ on wavelength. For stellar birth clouds, we adopt a simple power-law of index $-1.3$ appropriate for optically thick clouds with dust properties in the middle range of the Milky Way, the Large and the Small Magellanic Clouds \citep{dacunha2008}. In the ambient ISM, the shape of the effective absorption curve is more uncertain, as   the spatial distribution of dust relative to the stars can vary widely from galaxy to galaxy. Also, while \taulism\ represents an average optical depth along rays emanating in all directions from all stars, in practice, the attenuation will depend to some degree on the angle under which a galaxy is observed. Thus we write
\begin{equation}
\hat{\tau}_\lambda^{\,\mathrm{BC}}=(1-\mu)\hat{\tau}_V\left(\lambda/{5500\,\mathrm{\AA}}\right)^{-1.3}\,,
\label{eq:taulbc}
\end{equation}
\begin{equation}
\hat{\tau}_\lambda^{\,\mathrm{ISM}}=\mu \hat{\tau}_V\left(\lambda/{5500\,\mathrm{\AA}}\right)^{-n}\,.
\label{eq:taulism}
\end{equation}
In these expressions, $\tauv=\tauvbc+\tauvism$ is the total effective {\it V}-band absorption optical depth of the dust seen by young stars inside birth clouds, and $\mu=\tauvism/(\tauvbc+\tauvism)$ is the fraction of this contributed by dust in the ambient ISM. A value of $\mu$ around 0.3 and a typical slope $n=0.7$ in equation~(\ref{eq:taulism}) account remarkably well for several observed mean relations between the ultraviolet, optical and infrared properties of nearby galaxies (\citealt{charlot2000,kong2004,salim2007,dacunha2008}; Wild et al. 2011a\nocite{wild2011a}). More refined investigations using sophisticated dust models suggest that, in detail, the attenuation curve should flatten when the optical depth increases \citep{pierini2004,tuffs2004,rocha2008}. This is supported by the typically greyer optical attenuation curves observed in nearby edge-one galaxies compared to face-on ones (Wild et al. 2011b)\nocite{wild2011b}. A recent analysis shows that the trend favored by current models and observations can be reasonably well approximated by a simple expression of the slope of the attenuation curve of the form 
\begin{equation}
n\approx \frac{2}{1+\mu\tauv},
\label{eq:slope}
\end{equation}
with a scatter of about 20 percent at fixed $\mu\tauv$ (Chevallard et al., in preparation).
    
To compute $T^0_{\lambda}(t,t')$ in equations~(\ref{eq:lum})--(\ref{eq:tfunc}), we therefore proceed as follows. We first draw  \tauv\ between 0.01 and 4 from the probability density function $p(\tauv)=0.18\arctan[5(4-\tauv)]$, which is approximately uniform over the range from  $\tauv=0.01$ to $3$ and drops to zero at $\tauv=4$ (see Fig.~\ref{fig:dist}e below). We further draw $\mu$ from a uniform prior distribution between 0.1 and 0.7, and we draw $n$ uniformly over an interval $\Delta n=0.3$  centred on the mean value predicted by equation~({\ref{eq:slope}). Then, we report the selected \tauv, $\mu$ and $n$ in equations~(\ref{eq:trans0})--(\ref{eq:taulism}). This prescription accounts for a much broader range of plausible dust attenuation laws than the simple recipes usually adopted in galaxy spectral analyses \citep[e.g.,][]{kauffmann2003a, brinchmann2004,cid2005,panter2007,wuyts2009}. We therefore expect larger typical uncertainties arising from attenuation by dust than in these previous studies.


\subsection{Library of galaxy spectral energy distributions}
\label{sec:cats}

We use the library of star formation and chemical enrichment histories described in Section~\ref{sec:millennium} and the spectral models described in Section~\ref{sec:spectra} to simulate a large library of galaxy spectral energy distributions. This will be used in the remainder of this paper to assess the retrievability of galaxy physical parameters from various types of observations. To ensure that all kinds of spectral energy distributions are properly sampled, for each of the 500,000 star formation and chemical enrichment histories in the original library, we draw 10 different realizations of the evolutionary stage  (and hence of the parameter \fsfh) and, for star-forming galaxies, of the current physical properties (parameters \ssfr, \zgas, $\log\, U_0$, \xid, \tauv\ and $\mu$), as summarized in Table~\ref{tab:params}. The final library therefore contains 5 million models, which we can use to generate catalogs of galaxy spectra. 

\begin{table*}
\caption{Prior distributions of the current (i.e. averaged over the last 10\,Myr) physical properties of star-forming galaxies in the  library of star formation and chemical enrichment histories assembled in Sections~\ref{sec:millennium} and \ref{sec:spectra}.} 
\begin{center}
\begin{tabular}{l c c c}
\hline
Parameter & Description & Range & Probability Density Function \\
\hline
\ssfr & Specific star formation rate & $-2<\log(\ssfr/\mathrm{Gyr}^{-1})< 1$ & $0.05\left[1.5^5-\left|\log(\ssfr/\mathrm{Gyr}^{-1})+0.5\right|^5\right]$ \\
\zgas & Metallicity of the star-forming gas & $-1<\log(\zgas/\zsun)<0.54$ & uniform \\
$\log\,U_0$ & Zero-age ionization parameter &  $-3.5<\log\,U_0<-1.5$ & uniform \\
\xid & Dust-to-metal ratio in the ionized gas & $0.1<\xid<0.5$ & uniform \\
\tauv & Total $V$-band attenuation optical depth of the dust & $0.01<\tauv<4$ & $0.18\arctan\left[5(4-\tauv)\right]$\\
$\mu$ & Fraction of \tauv\ contributed by dust in diffuse ISM & $0.1<\mu<0.7$ & uniform \\
$n$ & Slope of the attenuation curve in the diffuse ISM & $1.6<n\,(1+\mu\tauv)<2.4$ & uniform \\
\hline
\end{tabular}
\end{center}
\label{tab:params}
\end{table*}

In the present study, we focus primarily on the rest-frame optical properties of galaxies. Our main goal is to compare the constraints that can be derived on galaxy physical parameters from different types of observations in this wavelength range. At wavelengths between 3525 and 7500\,\AA, the stellar population spectra computed using the models presented in Section~\ref{sec:stars} have a native spectral resolution of 2.3\,\AA\ (full width at half-maximum, hereafter FWHM), corresponding to a resolving power $R\approx2200$ at 5000\,\AA\ (the resolution is coarser outside this wavelength range). This allows us to investigate the retrievability of galaxy physical parameters from 4 different types of observations at optical wavelengths:
\begin{itemize}
\item {\it Medium-resolution spectroscopy ($\lambda$=3600--7400\,\AA).} We adopt a fiducial resolution of 5\,\AA\  FWHM, corresponding to $R=1000$ at $\lambda=5000\,$\AA, and a fixed pixel size of 2.5\,\AA.
\item {\it Low-resolution spectroscopy ($\lambda$=3600--7400\,\AA).} We adopt a fiducial resolution of 50\,\AA\  FWHM, corresponding to $R=100$ at $\lambda=5000\,$\AA, and a fixed pixel size of 25\,\AA.
\item {\it Equivalent-width measurements of strong emission lines.} We also investigate the constraints set purely by the equivalent widths of the following emission lines, at both low and medium spectral resolution: $\oii\lambda3727$; \hb; $\oiii\lambda4959$, $\oiii\lambda5007$; $\nii\lambda6548$; \ha; $\nii\lambda6584$; $\sii\lambda6716$ and $\sii\lambda6731$. An advantage of equivalent widths is that they can provide useful information even in spectra which are not flux-calibrated.
\item {\it Broadband photometry.} We adopt the SDSS $ugriz$ filter response functions.
\end{itemize}
As an example, Fig.~\ref{fig:methods} shows the spectral energy distribution of the same model galaxy as in Fig.~\ref{fig:sfhmfh} at the age of 8.4\,Gyr (with current physical parameters resampled using the distributions in Table~\ref{tab:params}), computed  alternatively at medium (panel a) and low (panel b) spectral resolution and convolved with the SDSS  $ugriz$ filter response functions (panel c).

\begin{figure}
\begin{center}
\includegraphics[width=0.50\textwidth]{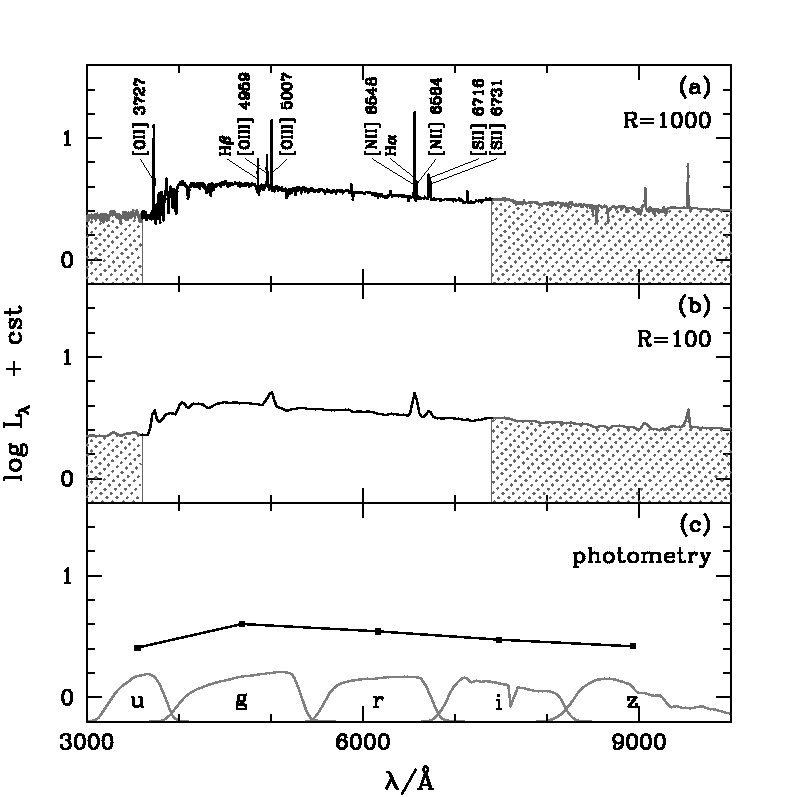}
\caption{Spectral energy distribution of the same model galaxy as in Fig.~\ref{fig:sfhmfh} at the age of 8.4\,Gyr, computed using the models described in Sections~\ref{sec:millennium} and \ref{sec:spectra}. ({\it a}) At a spectral resolution of 5\,\AA\  FWHM ($R=1000$ at 5000\,\AA). The black portion of the spectrum ($\lambda$=3600--7400\,\AA) is that used to retrieve galaxy physical parameters at this resolution. ({\it b}) Same as in ({\it a}), but at a spectral resolution of 50\,\AA\  FWHM ($R=100$ at 5000\,\AA). ({\it c}) Same as in ({\it a}), but convolved with the SDSS $ugriz$ filter response functions (shown at the bottom). The model galaxy has $\fsfh=0.3$ and current parameters (resampled using the distributions in Table~\ref{tab:params}) $\ssfr=0.08\,\mathrm{Gyr}^{-1}$, $\zgas=0.4\zsun$, $\log U_0=-2.8$, $\xid=0.3$, $\tauv=1.0$, $\mu=0.6$ and $n=0.7$. In ({\it a}) and ({\it b}), the adopted pixel size is half the size of a resolution element.}
\label{fig:methods}
\end{center}
\end{figure}

\subsection{Retrievability of galaxy physical parameters}
\label{sec:stat}

The galaxy spectral library generated in the previous section can be used to assess the retrievability of physical parameters from different types of observations. To achieve this, we first draw a random subsample of 10,000 spectral energy distributions from the library, to which we apply artificial noise to mimic true observations. For low- and medium-resolution spectroscopy, we apply noise independent of wavelength (i.e. white noise), parametrized in terms of the median \sn\ per pixel, \snmed. We consider the cases $\snmed=5$ and 20 (for reference, the stellar population spectra computed using the models presented in Section~\ref{sec:stars} have a native $\snmed$ of about 95). For multiband photometry, instead, we take the signal-to-noise ratio to be the same in all wavebands ($\sn=10$ or 30). In the following, we refer to this subsample of 10,000 noisy spectral energy distributions as `pseudo-observations', and to the 4,990,000 remaining spectral energy distributions in the library as simply `models'.
 
We use a Bayesian approach to quantify the accuracy to which several physical parameters of these pseudo-galaxies can be retrieved from different types of observables. The likelihood of the $j\,$th model, noted ${\cal M}_j$, given the spectral energy distribution of a pseudo-galaxy  can be written
\begin{equation}
\label{eq:prob}
\ln P({\cal M}_j|\{{\cal O}_i\})=-\frac{1}{2}\sum_i \left ( \frac{{\cal O}_i-w_j{\cal M}_{j,i}}{\sigma_i} \right )^{2}\,,
\end{equation}
where ${\cal O}_i$ and ${\cal M}_{j,i}$ are the spectral energy distributions (i.e., either the spectral flux density at the $i\,$th wavelength, or the equivalent width of the $i\,$th emission line, or the flux in the $i\,$th photometric band; see Section~\ref{sec:cats}) of the pseudo-galaxy and the $j\,$th model, respectively, $\sigma_i$  is the uncertainty in ${\cal O}_i$ corresponding to the adopted signal-to-noise ratio, and
\begin{equation}
w_j=\left ( \sum_i \frac{{\cal O}_i{\cal M}_{j,i}}{\sigma_i^2} \right ) \left[\sum_i \left(\frac{{\cal M}_{j,i}}{\sigma_i}\right )^2\right]^{-1}
\label{eq:weight}
\end{equation}
is the model scaling factor that maximizes $P({\cal M}_j|\{{\cal O}_i\})$ (this parameter is unity in the case of scale-free equivalent widths). In practice, in the case of medium-resolution spectroscopy, the summation on the right-hand side of expression~(\ref{eq:prob}) extends over 1520 wavelength points and hence generally yields values greater than 2000, even for a good-fit model.\footnote{We note that only a small contribution to these high values arises from our neglect of intrinsic correlations between spectral pixels in equation~(\ref{eq:prob}). A detailed investigation reveals that such correlations are significant only for pixels associated with emission lines, which reflects the fact that all emission lines arise from the same ionizing radiation. We have used a test sample of 3000 pseudo-galaxies to check the effect of accounting for pixel-to-pixel correlations on the calculation of model likelihoods. This requires the introduction of the covariance matrix in equation~(\ref{eq:prob}) and is computationally very demanding. The result is a negligible effect on the computed model likelihoods. Hence  equation~(\ref{eq:prob}) should be appropriate to compute model likelihoods.} The absolute likelihood in this case cannot be evaluated numerically, as it is always rounded to 0. For medium-resolution spectroscopy, therefore, we rather compute the probability of the $j\,$th model relative to the best-fitting model, noted ${\cal M}_0$,
\begin{equation}
\label{eq:relprob}
\ln \tilde{P}({\cal M}_j|\{{\cal O}_i\})=
\ln P({\cal M}_j|\{{\cal O}_i\})-\ln P({\cal M}_0|\{{\cal O}_i\})\,.
\end{equation}
We have checked that in the case of low-resolution spectroscopy, where both $P({\cal M}_j|\{{\cal O}_i\})$ and $\tilde{P}({\cal M}_j|\{{\cal O}_i\})$ can be evaluated numerically, the likelihood distributions of galaxy physical parameters obtained using equations~(\ref{eq:prob}) and (\ref{eq:relprob}) are similar.

In this paper, we focus on the retrievability of 6 physical parameters from the observed spectral energy distributions of galaxies:
\begin{itemize}
\item the observer-frame absolute $r$-band (stellar) mass-to-light ratio, \mlr. This is the ratio of the galaxy stellar mass to the absolute luminosity in the observed $r$ band;
\item the fraction of the current galaxy stellar mass formed during the last 2.5\,Gyr, \fsfh;
\item the specific star formation rate, \ssfr;
\item the gas-phase oxygen abundance in units of \aboh, where \oh\ is the abundance by number of oxygen relative to hydrogen. In the models of nebular emission adopted in Section~\ref{sec:neb}, the gas-phase oxygen abundance corresponding to solar metallicity ($\zgas=\zsun$) is 8.81, 8.73 and 8.63  for dust-to-metal ratios $\xid=0.1$, 0.3 and 0.5, respectively.
\item the total effective $V$-band absorption optical depth of the dust, \tauv\ (Section~\ref{sec:dust});
\item the fraction of \tauv\ arising from dust in the ambient ISM, $\mu$.
\end{itemize}

Fig.~\ref{fig:dist} shows the prior distributions of these parameters in the galaxy spectral library generated in Section~\ref{sec:cats}. In each panel, the shaded histogram shows the distribution including all galaxies, while the solid histogram shows the contribution by star-forming galaxies alone. Galaxies without star formation have typically high $\mlr$  in Fig.~\ref{fig:dist}a and $\fsfh\approx0$ in Fig.~\ref{fig:dist}b. These galaxies are off scale (at $\log\ssfr=-\infty$) in Fig.~\ref{fig:dist}c. They do not contribute to the histograms in Figs~\ref{fig:dist}d--f, which pertain to interstellar parameters. As emphasized in Sections~\ref{sec:millennium} and \ref{sec:spectra}, the prior distributions of Fig.~\ref{fig:dist} are as flat as possible to avoid unwanted biases in the retrieval of physical parameters from observed spectral energy distributions of galaxies. This choice is motivated by the desire to make the spectral library appropriate for the analysis of galaxies in wide ranges of physical properties (we note that the finite extent of the prior distributions in Fig.~\ref{fig:dist} could still bias analyses of galaxies with extreme parameters). Alternative prior distributions can be designed  to help reduce the uncertainties in the parameters derived for galaxies with known specific physical properties (e.g., actively star-forming, early-type, metal-poor). We return to the influence of prior distributions on parameter determinations in Section~\ref{sec:prior} below, where we investigate the results obtained using the library of galaxy star formation and chemical enrichment histories derived from the original semi-analytic recipes of \citet{delucia2007} .

\begin{figure}
\begin{center}
\includegraphics[width=0.47\textwidth]{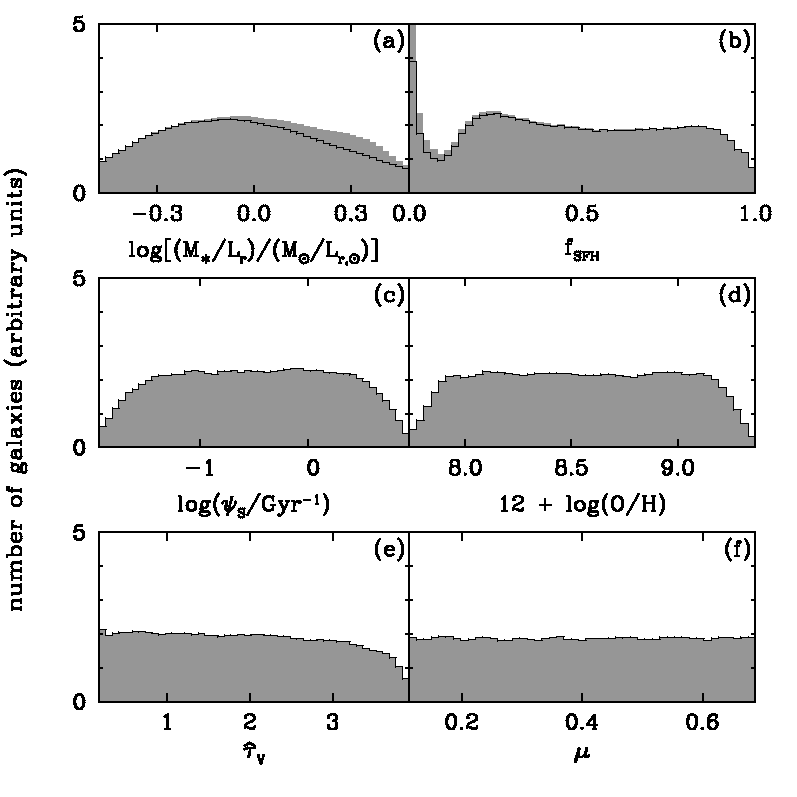}
\caption{Prior distributions of selected physical parameters of the 5 million galaxies in the spectral library generated in Section~\ref{sec:cats}: ({\it a}) observer-frame absolute $r$-band stellar mass-to-light ratio, \mlr; ({\it b}) fraction of current stellar mass formed during the last 2.5\,Gyr, \fsfh; ({\it c}) specific star formation rate, \ssfr; ({\it d}) gas-phase oxygen abundance, $\aboh$; ({\it e}) total effective $V$-band absorption optical depth of the dust, \tauv; ({\it f}) fraction of \tauv\ arising from dust in the ambient ISM, $\mu$. In each panel, the shaded histogram shows the distribution for all galaxies, while the solid histogram shows the contribution by star-forming galaxies alone. Non-star-forming galaxies are off scale (at $\log\ssfr=-\infty$) in panel ({\it c}) and do not contribute to the distributions of interstellar parameters in panels ({\it d})--({\it f}).}
\label{fig:dist}
\end{center}
\end{figure}

\begin{figure}
\begin{center}
\includegraphics[width=0.50\textwidth]{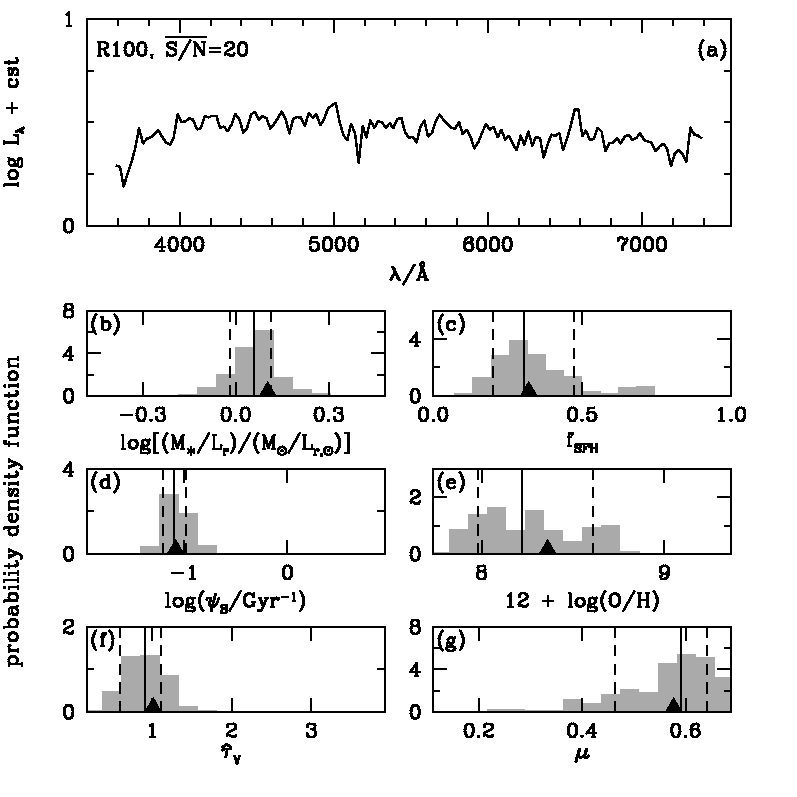}
\caption{Probability density functions of the same physical parameters as in Fig.~\ref{fig:dist} retrieved, using the Bayesian approach described in Section~\ref{sec:stat}, from the pseudo-galaxy spectrum shown at the top. The spectrum was obtained by adding artificial noise with $\snmed=20$ to the $R=100$ model spectrum of Fig.~\ref{fig:methods}b. In each panel, the black triangle indicates the true value of the parameter of the pseudo-galaxy, the solid line the best estimate (50th percentile of the retrieved likelihood distribution) and the dashed lines the associated 68-percent confidence interval (16th and 84th percentiles).}
\label{fig:pdf}
\end{center}
\end{figure}

In Fig.~\ref{fig:pdf}, we show examples of the probability density functions (likelihood distributions) of physical parameters retrieved, using equations~(\ref{eq:prob})--(\ref{eq:weight}) and the prior distributions of Fig.~\ref{fig:dist}, from the low-resolution, high-\sn\ spectrum of a pseudo-galaxy in our library. This pseudo-observation was obtained by adding artificial noise ($\snmed=20$) to the model spectrum of Fig.~\ref{fig:methods}b. We take the `best estimate' of each parameter in Fig.~\ref{fig:pdf} to be the median of the retrieved probability density function and the associated `uncertainty' to be half the 16th--84th percentile range (this would be equivalent to the $\pm1\sigma$ range in the case of a Gaussian distribution). To quantify potential biases in the retrieved likelihood distribution, we further define the `accuracy' as the absolute difference between the retrieved best estimate and true parameter value. In this example, all the parameters are well retrieved. The success in recovering the fraction \fsfh\ of intermediate-age stars (Fig.~\ref{fig:pdf}c) and the interstellar parameters \aboh, \tauv\  and $\mu$ (Figs~\ref{fig:pdf}e--\ref{fig:pdf}g) is remarkable. It is the first illustration of the power of our approach to exploit strong absorption- and emission-line signatures in noisy, low-resolution spectra. We examine this property in detail in Section~\ref{sec:spec} below.

It is important to stress that the approach adopted here to assess the retrievability of galaxy physical parameters from different types of observations, while idealized, is as realistic as can be achieved using modern techniques. Pseudo-observations are required to test the retrievability of physical parameters from observables, because the physical parameters of observed galaxies cannot be known a priori. The fact that we draw pseudo-observations from the same spectral library as the models used to analyze them makes the constraints derived on the retrievability of physical parameters optimistic: our results correspond to the best possible case in which the models are a realistic representation of true galaxies. In Section~\ref{sec:sample} below, we also illustrate the constraints that can be derived by applying our approach to interpret a sample of observed galaxy spectra.


\section{Constraints on galaxy physical parameters from different types of observations}
\label{sec:estim}

In this section, we use the modeling approach described in the previous section to quantify the accuracy to which physical parameters of galaxies can be retrieved from different types of observations. We focus on the retrievability of the 6 parameters introduced in Section~\ref{sec:stat}: the observer-frame absolute $r$-band stellar mass-to-light ratio, \mlr; the fraction of the current galaxy stellar mass formed during the last 2.5\,Gyr, \fsfh; the specific star formation rate, \ssfr; the gas-phase oxygen abundance, $\aboh$; the total effective $V$-band absorption optical depth of the dust, \tauv; and the fraction of \tauv\ arising from dust in the ambient ISM, $\mu$. In Section~\ref{sec:phot} below, we first examine the constraints set on these parameters by multi-band photometric observations. We consider the influence of several factors, such as the signal-to-noise ratio, the number of photometric bands, the contamination of broad-band fluxes by nebular emission lines and the uncertainty introduced by the lack of redshift information. We find that photometric observations can provide useful constraints mainly on \mlr\ and \ssfr. Then, in Section \ref{sec:spec}, we quantify the improvement that can be achieved in the constraints on all parameters by appealing to different types of spectroscopic observations.

\subsection{Constraints from multi-band photometry}
\label{sec:phot}

\subsubsection{Nearby galaxies}
\label{sec:nearphot}

We first explore the extent to which physical parameters of nearby galaxies can be constrained using 5-band $ugriz$ photometry. We extract 10,000 spectral energy distributions from the library assembled in Section~\ref{sec:cats} to compute pseudo-observations of galaxies in these bands, as described in Section~\ref{sec:stat}. We adopt for the moment a signal-to-noise ratio $\sn=30$ in all bands and place all spectral energy distributions at $z=0$. The rest of the 5 million models in the library allow us to retrieve likelihood distributions of \mlr, \fsfh, \ssfr, \aboh, \tauv\ and $\mu$ for each of these 10,000 pseudo-galaxies, by means of equations~(\ref{eq:prob})--(\ref{eq:weight}). In the following, we refer to this setting as the `standard' case.

\begin{figure*}
\begin{center}
\includegraphics[width=\textwidth]{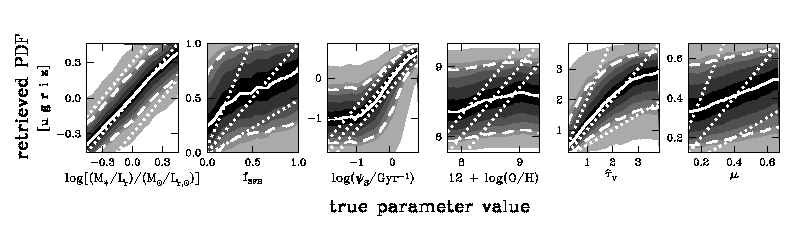}
\caption{Average probability density functions of the same 6 physical parameters as in Fig.~\ref{fig:dist} retrieved, using the Bayesian approach described in Section~\ref{sec:stat}, from 5-band $ugriz$ photometry with $\sn=30$ of a sample of 10,000 pseudo-galaxies (standard case). Each panel corresponds to a different parameter [from left to right: \mlr, \fsfh, \ssfr, \aboh, \tauv\ and $\mu$]. In each case, average probability density functions in 50 narrow bins of true parameter value were obtained by coadding and then renormalizing the probability density functions retrieved for the 10,000 pseudo-galaxies (see Section~\ref{sec:nearphot} for detail). Grey levels locate the 2.5th, 12th, 16th, 22nd, 30th, 40th, 60th, 70th, 78th, 84th, 88th and 97.5th percentiles of the average likelihood distribution in each bin. The  solid line locates the associated median (best estimate) and the 2 dashed lines the 16th and 84th percentiles (68-percent confidence interval). For reference, 3 dotted lines indicate the identity relation and deviations by a factor of 2 ($\pm0.3\,$dex) between the retrieved and true parameter values.}
\label{fig:photbase}
\end{center}
\end{figure*}

Fig.~\ref{fig:photbase} summarizes the constraints that can be set on the physical parameters of nearby galaxies using high-quality, 5-band $ugriz$ photometry. For each parameter, we plot the average probability density function retrieved from the above analysis as a function of true parameter value. This average was obtained by coadding and then renormalizing the probability density functions retrieved in narrow bins of true parameter value for the 10,000 pseudo-galaxies in our sample. The solid line in each panel of Fig.~\ref{fig:photbase} locates the median (i.e. best estimate; Section~\ref{sec:stat}) of the average likelihood distribution in each bin, the dashed lines the 16th and 84th percentiles (68-percent confidence interval) and the outer edges of the lightest grey level the 2.5th and 97.5th percentiles (these would correspond to the $\pm2\sigma$ levels in the case of a Gaussian distribution). For reference, 3 dotted lines indicate the identity relation and deviations by a factor of 2 ($\pm0.3\,$dex) between the retrieved and true parameters.

Fig.~\ref{fig:photbase} shows that the $r$-band stellar mass-to-light ratio of nearby galaxies can be reasonably well constrained using 5-band $ugriz$ photometry. The retrieved best estimate is in excellent agreement with the true parameter value over most of the explored range, and the associated uncertainty, defined as half the 16th--84th percentile range (Section~\ref{sec:stat}), is only about 0.19\,dex (this number would be larger if we allowed for IMF variations).  At the extremities of the  range, the relative deficiency of galaxies with very small and very large \mlr\ in the spectral library (Fig.~\ref{fig:dist}a) slightly biases the retrieved likelihood distributions toward intermediate \mlr\ values. Fig.~\ref{fig:photbase} also shows that, in contrast to \mlr, the fraction \fsfh\ of stellar mass formed in the past 2.5\,Gyr cannot be well retrieved from 5-band $ugriz$ photometry. This is because the strong absorption features of intermediate-age stars have a negligible influence on broadband fluxes. The signatures of \fsfh\ on broadband colours are otherwise difficult to disentangle from those of stellar metallicity and attenuation by dust.

The constraints on specific star formation rate from 5-band $ugriz$ photometry in Fig.~\ref{fig:photbase} are meaningful only for actively star-forming galaxies, i.e. for $\ssfr>1\,\textrm{Gyr}^{-1}$, and the associated uncertainty is at best 0.30\,dex. In such galaxies, the contamination of broadband fluxes by strong emission lines produces identifiable signatures (\oii\ in the $u$ band; \hb\ and \oiii\ in the $g$ band; \ha, \nii\ and \sii\ in the $r$ band). Even then, the similar colours of galaxies with both low \ssfr\ and \tauv\ and those with both high \ssfr\ and \tauv\ give rise to long tails toward low \ssfr\ in the retrieved likelihood distributions. This degeneracy is illustrated by the overlap of galaxies with different \ssfr\ in Fig~\ref{fig:color}a, which shows the $g-i$ versus $u-g$ colours of a subset of 50,000 models in the spectral library. Together with the fact that, even in galaxies with strong emission lines, broadband colours do not allow the distinction between different line ratios, this also implies poor constraints on the gas-phase oxygen abundance and the dust parameters \tauv\ and $\mu$ in Fig.~\ref{fig:photbase}. The retrieved likelihood distributions of these parameters are largely driven by the prior distributions of Fig.~\ref{fig:dist}.

\begin{figure}
\begin{center}
\includegraphics[width=0.50\textwidth]{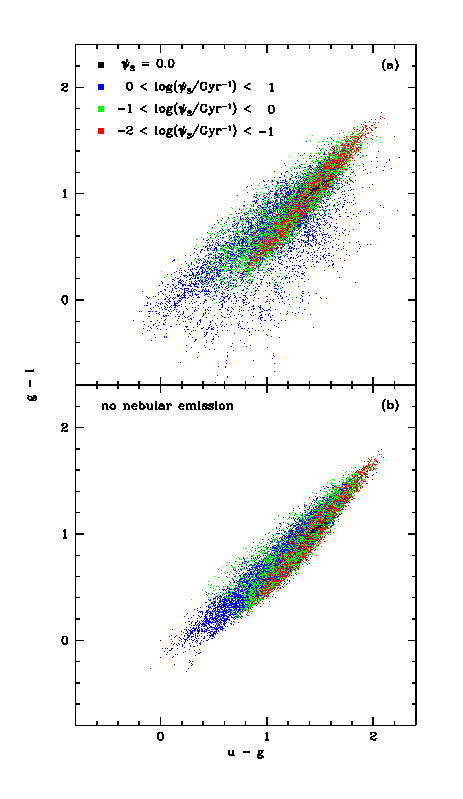}
\caption{({\it a}) $g-i$ colour plotted against $u-g$ colour for a subset of 50,000 models from the galaxy spectral library assembled in Section~\ref{sec:cats}. The models are colour-coded according to specific star formation rate, as indicated. ({\it b}) Same as ({\it a}), but without including the contribution by nebular emission to broadband fluxes.}
\label{fig:color}
\end{center}
\end{figure}

The results of Fig.~\ref{fig:photbase} are consistent with those of \citet{wuyts2009}, who explored the recovery of physical parameters from broadband spectral energy distributions of model galaxies taken from hydrodynamical merger simulations. Our study differs from theirs in that we consider much wider ranges of star formation and chemical enrichment histories and dust properties to retrieve likelihood distributions of physical parameters. Also, \citet{wuyts2009} neglect the contribution by nebular emission to broadband fluxes and include 11 bands to sample the rest-frame spectral energy distributions of galaxies over a wide range of wavelengths, between roughly 0.15 and 3\,$\mu$m. Finally, they analyze galaxy simulations in the redshift range $1.5\leq z\leq2.9$, which span a much narrower range of stellar population ages than the pseudo-galaxies in our sample. For all these reasons, \citet{wuyts2009} find smaller typical uncertainties in the retrieved stellar mass ($\sim0.10\,$dex) and star formation rate ($\sim0.31\,$dex) than can be inferred from the constraints on \mlr\ and \ssfr\ in Fig.~\ref{fig:photbase}. They also find much tighter constraints on the dust attenuation optical depth ($\sim0.31$). \citet{wuyts2009} note that using simple (single-burst, constant and exponentially declining) star formation histories to fit the results of more realistic hydrodynamical merger simulations can induce significant deviations between the retrieved maximum-likelihood and true parameters values. Our adoption of the sophisticated galaxy spectral library assembled in Section~\ref{sec:cats} should minimize this type of limitation when applying our approach to the interpretation of real observations (Section~\ref{sec:sample}).

We now return to our investigation and examine the sensitivity of the results of Fig.~\ref{fig:photbase} to the adopted signal-to-noise ratio, the photometric wavelength coverage and the inclusion of nebular emission when interpreting broadband observations of galaxies. To quantify potential differences in the retrieved likelihood distributions relative to the standard case, we define the improvement factor as the ratio
\begin{equation}
\label{eq:isig}
\isig={{\textrm{uncertainty in the standard case}}\over{\textrm{uncertainty}}}\,.
\end{equation}
This quantity can be computed for any true parameter value. It exceeds unity when the constraints are tighter (i.e. the uncertainty smaller) than in the standard case. We further define the gain in accuracy as the difference
\begin{equation}
\label{eq:delta}
\Delta=(\textrm{accuracy in the standard case})-(\textrm{accuracy})\,.
\end{equation}
This quantity is positive when the retrieved best estimate is more accurate (i.e. the absolute difference between the retrieved best estimate and true parameter value smaller) than in the standard case. In Fig.~\ref{fig:photdeg}, we show the analog of Fig.~\ref{fig:photbase} for 3 distinct alternatives to the standard case: adopting a signal-to-noise ratio $\sn=10$ instead of 30 (top row); removing the constraints from the reddest 2 photometric bands, $i$ and $z$ (middle row); and not including nebular emission in the model library used to analyze pseudo-observations (bottom row). In all cases, we show at the bottom of each panel the improvement factor \isig\ and the gain in accuracy $\Delta$ as a function of true parameter value.

Fig.~\ref{fig:photdeg} (top row) shows that reducing the signal-to-noise ratio from $\sn=30$ to 10 makes the uncertainty in the retrieved \mlr\ larger by about 15 percent and, at the extremities of the explored range, the accuracy between 10 and 25 percent worse. A more dramatic difference relative to the standard case arises from the loss of the emission-line signal for actively star-forming galaxies. This causes a sharp drop in \isig\  at high \ssfr, implying poor constraints on the specific star formation rate over the entire explored range. Interestingly, Fig.~\ref{fig:photdeg} (middle row) shows that the effect of removing the $i$- and $z$-band constraints at $\sn=30$ is very similar to that of lowering the signal-to-noise ratio of 5-band $ugriz$ photometry. This is because the signatures of dust attenuation can be less well delineated from those of star formation from the {\it ugr} colours alone, even if the contamination of these broadband fluxes by strong emission lines can in principle be detected at high \sn\  (Fig.~\ref{fig:photbase}). As expected, the other parameters, \fsfh, \aboh, \tauv\ and $\mu$, can be even less well retrieved than in the standard case when the signal-to-noise ratio is lowered and when the $i$- and $z$-band constraints are removed.

It is also of interest to quantify the importance of including nebular emission when interpreting broadband observations of galaxies. A comparison between the top and bottom panels of Fig.~\ref{fig:color} shows that the contribution by \oii\ emission to the $u$-band flux and that of  \hb\ and \oiii\ emission to the $g$-band flux can change the broadband $u-g$ and $g-i$ colours of  galaxies with $\ssfr>1\,\textrm{Gyr}^{-1}$ by several tenths of a magnitude. In Fig.~\ref{fig:photdeg} (bottom row), we show the results of attempting to interpret the same pseudo-observations as in Fig.~\ref{fig:photbase} (which include the contribution by nebular emission to the $ugriz$ fluxes) using a spectral library which neglects nebular emission. This spectral library was constructed in the same way as that assembled in Section~\ref{sec:cats}, but ignoring the prescription for  nebular emission in Section~\ref{sec:neb}. The retrieved \mlr\ in this case is very similar to that obtained in the standard case, except for the slightly worse accuracy ($\Delta<0$) at very low \mlr, consistent with the contamination of the $r$-band flux by \ha, \nii\ and \sii\ emission in actively star-forming galaxies. As anticipated from Fig.~\ref{fig:color}, models without nebular emission do not sample broadband colours well enough to allow the retrieval of meaningful constraints on \ssfr\  for actively star-forming galaxies. The other parameters, \fsfh, \aboh, \tauv\ and $\mu$, are as poorly constrained as in the standard case. We conclude from Fig.~\ref{fig:photdeg} that high \sn, good wavelength coverage and accounting for nebular emission are all important to tighten the constraints on specific star formation rate from broadband observations of galaxies, while the inclusion of nebular emission is less crucial to constrain stellar mass-to-light ratio.

\begin{figure*}
\begin{center}
\includegraphics[width=\textwidth]{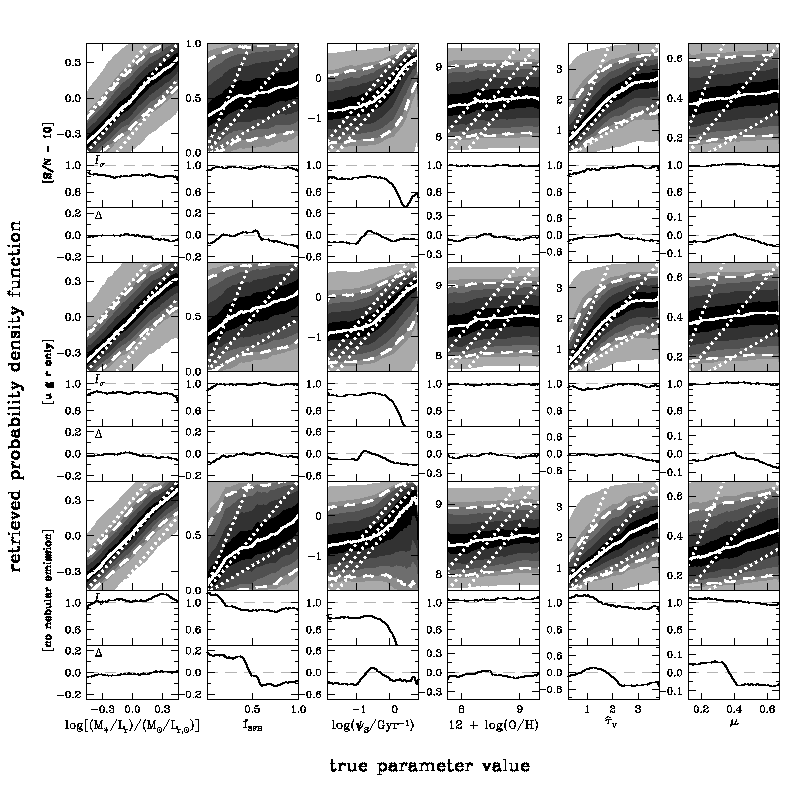}
\caption{Same as Fig.~\ref{fig:photbase}, but for 3 distinct alternatives to the standard case: ({\it top row}) adopting a signal-to-noise ratio $\sn=10$ instead of 30; ({\it middle row}) using constraints from only the $ugr$ photometric bands instead of $ugriz$; and ({\it bottom row}) not including nebular emission in the model library used to analyze the sample of 10,000 pseudo-galaxies. In all cases, the improvement factor \isig\ and the gain in accuracy $\Delta$ (equations~\ref{eq:isig}--\ref{eq:delta}) are shown as a function of true parameter value at the bottom of each panel to quantify differences in the retrieved likelihood distributions relative to the standard case of Fig.~\ref{fig:photbase}.}
\label{fig:photdeg}
\end{center}
\end{figure*}

\subsubsection{Galaxies at unknown redshift}
\label{sec:zphot}

We have tested so far the retrievability of physical parameters from 5-band $ugriz$ photometry only for nearby galaxies. We now wish to explore the constraints achievable for galaxies at unknown redshift. \citet{ilbert2006} have shown that redshift can be best estimated from 5-band $ugriz$ photometry out to $z\sim1.5$. Photometric redshift measurements of more distant galaxies require observations at longer wavelengths \citep[e.g.][]{ilbert2009}, which are beyond the scope of the present paper (see Pacifici et al., in preparation). We therefore restrict ourselves here to the retrievability of physical parameters of galaxies from 5-band $ugriz$ photometry at redshifts in the range $0<z<1$.

We use an approach similar to that described in Section~\ref{sec:cats} to generate a new library of 5 million spectral energy distributions of galaxies at random redshifts between 0 and 1. For each of the 500,000 star formation and chemical enrichment histories in the original library of Section~\ref{sec:millennium}, we first draw an observation redshift \zobs\ between 0 and 1. Then, we proceed as outlined in Section~\ref{sec:cats} and draw 10 different realization of the evolutionary stage (i.e. galaxy age) at which the galaxy is looked at (uniformly in redshift between \zobs\ and $z=1.7$, to better sample the parameter \fsfh; see Section~\ref{sec:millennium}) and, for star-forming galaxies, of the current parameters \ssfr, \zgas, $\log\, U_0$, \xid, \tauv\ and $\mu$ (using the distributions in Table~\ref{tab:params}). The resulting spectral library contains about 8 percent early-type galaxies, i.e. slightly less than the 10 percent in the $z=0$ library assembled in Section~\ref{sec:cats}. Fig.~\ref{fig:disthighz} shows the prior distributions of the physical parameters \mlr, \fsfh, \ssfr, \aboh, \tauv\ and $\mu$ in this library. As in Fig.~\ref{fig:dist}, in each panel, the shaded histogram shows the distribution including all galaxies, while the solid histogram shows the contribution by star-forming galaxies alone. The prior distribution of the observer-frame absolute $r$-band stellar mass-to-light ratio in Fig.~\ref{fig:disthighz}a reaches larger values than in Fig.~\ref{fig:dist}a because of the contribution by (dusty) high-redshift galaxies with large K corrections. The early  evolutionary stage of high-redshift galaxies also implies a larger typical \fsfh\ in Fig.~\ref{fig:disthighz}b than in Fig.~\ref{fig:dist}b. By construction, the distributions of the current physical parameters \ssfr, \aboh, \tauv\ and $\mu$ are the same in Figs~\ref{fig:dist} and \ref{fig:disthighz} (from Table~\ref{tab:params}).

\begin{figure}
\begin{center}
\includegraphics[width=0.45\textwidth]{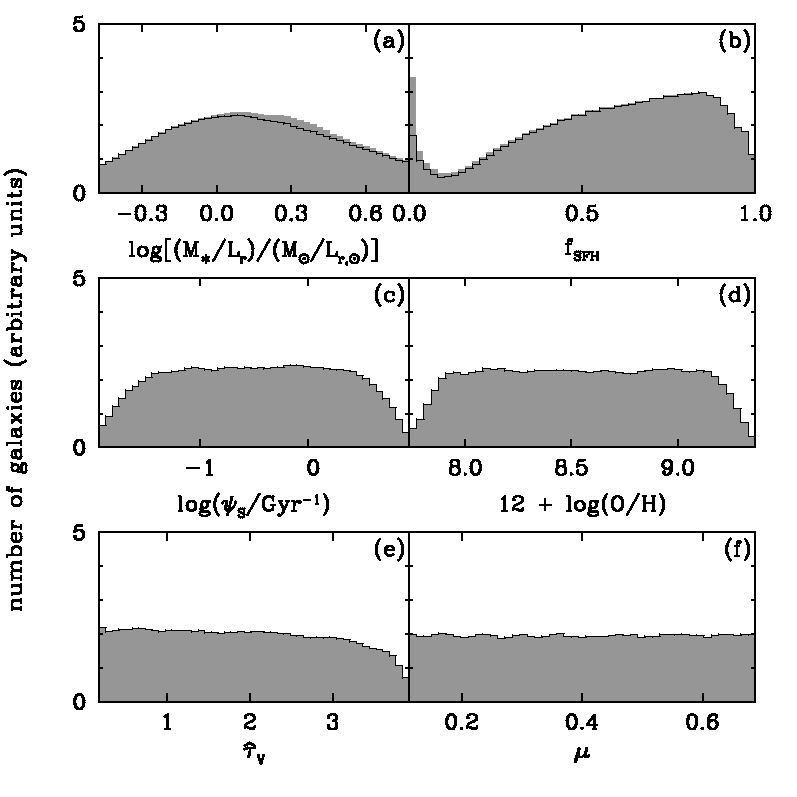}
\caption{Prior distributions of the same physical parameters as in Fig.~\ref{fig:dist} for the 5 million galaxies at redshifts between 0 and 1 in the spectral library assembled in Section~\ref{sec:zphot}. In each panel, the shaded histogram shows the distribution for all galaxies, while the solid histogram shows the contribution by star-forming galaxies alone. Non-star-forming galaxies are off scale (at $\log\ssfr=-\infty$) in panel ({\it c}) and do not contribute to the distributions of interstellar parameters in panels ({\it d})--({\it f}).}
\label{fig:disthighz}
\end{center}
\end{figure}

By analogy with Section~\ref{sec:nearphot}, we generate pseudo-observations in the $ugriz$ photometric bands of 10,000 galaxies in the new library. These pseudo-observations sample different wavelength ranges of the rest-frame spectral energy distributions of galaxies at different redshifts, from roughly 3000--10,000 \AA\ at $\zobs=0$ to 1500--5000 \AA\ at $\zobs=1$. We consider both high-quality ($\sn=30$) and low-quality ($\sn=10$) observations. Using the rest of the 5 million models in the library, we can retrieve likelihood distributions of \zobs\ for each of the 10,000 pseudo-galaxies, as well as likelihood distributions of  \mlr, \fsfh, \ssfr, \aboh, \tauv\ and $\mu$. 

In Fig.~\ref{fig:phothisto}a, we plot the distribution of the difference between the retrieved best estimate (i.e. median) and true value of \zobs, in units of $1+\zobs$, for the 10,000 pseudo-galaxies in our sample. The results for $\sn=30$ (shaded histogram) are in remarkable agreement with the $\sigma=0.03$ Gaussian uncertainty in this quantity quoted by \citet{ilbert2006}, who analyzed CFHTLS ($\sn\ga30$) $u^*g'r'i'z'$ photometry of about 3000  galaxies with spectroscopic redshifts between 0.2 and 1.5 (dashed line). For $\sn=10$, the retrieved best estimate of \zobs\ in Fig.~\ref{fig:phothisto}a remains in good general agreement with the true value (solid histogram). To better illustrate the origin of the histograms in Figs~\ref{fig:phothisto}a, we show in Figs~\ref{fig:phothisto}b and \ref{fig:phothisto}c the detail of the average retrieved probability density functions as a function of \zobs\ for $\sn=30$ and 10, respectively.

\begin{figure}
\begin{center}
\includegraphics[width=0.45\textwidth]{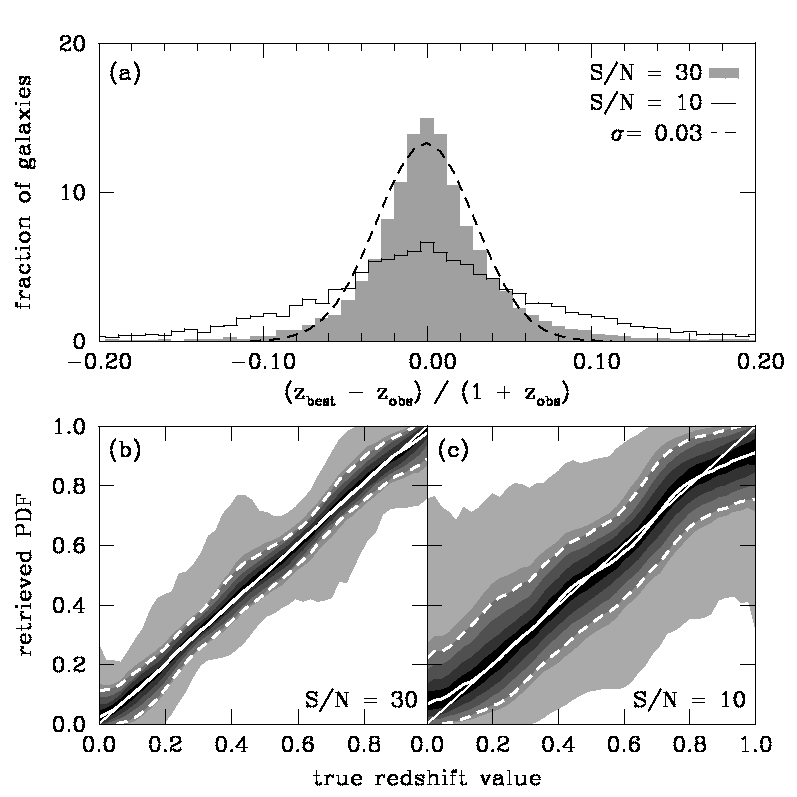}
\caption{({\it a}) Distribution of the difference between the retrieved best estimate (i.e. median) and true value of \zobs, in units of $1+\zobs$, for a sample of 10,000 pseudo-galaxies at redshifts between 0 and 1 observed in the $ugriz$ photometric bands. The shaded and solid histograms show the distributions obtained for $\sn=30$ and $\sn=10$, respectively. For reference, a dashed line indicates the $\sigma=0.03$ Gaussian accuracy quoted by \citet{ilbert2006}, who analyzed CFHTLS ($\sn\ga30$) $u^*g'r'i'z'$ photometry of about 3000  galaxies with spectroscopic redshifts between 0.2 and 1.5. ({\it b}) Detail of the average retrieved probability density function plotted against \zobs, for $\sn=30$. The lines and shading have the same meaning as in Fig.~\ref{fig:photbase}. ({\it c}) Same as ({\it b}), but for $\sn=10$.}
\label{fig:phothisto}
\end{center}
\end{figure}

\begin{figure*}
\begin{center}
\includegraphics[width=\textwidth]{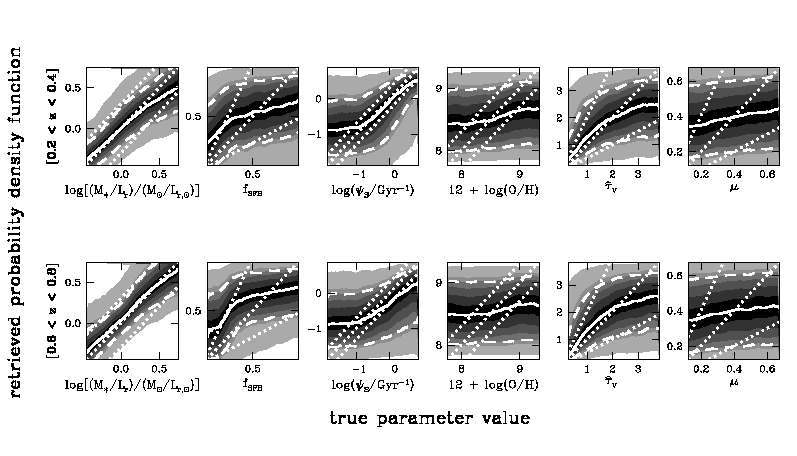}
\caption{Average probability density functions of the same 6 physical parameters as in Fig.~\ref{fig:disthighz} retrieved, using the Bayesian approach described in Section~\ref{sec:stat}, from 5-band $ugriz$ photometry with $\sn=30$ of samples of 10,000 pseudo-galaxies at random redshifts: ({\it top row}) drawn in the redshift range $0.2<\zobs<0.4$; and ({\it bottom row}) drawn in the redshift range $0.6<\zobs<0.8$. In each case, the redshift \zobs\ of a galaxy is assumed not to be known a priori, and the probability density functions are computed using models at all redshifts between 0 and 1 in the spectral library assembled in Section~\ref{sec:zphot}. In all panels, the lines and shading have the same meaning as in Fig.~\ref{fig:photbase}.}
\label{fig:photzzz}
\end{center}
\end{figure*}

The retrieved probability density functions of the other physical parameters of interest to us are difficult to interpret when averaged over galaxies at all redshifts between 0 and 1, because of the different effects impacting the recovery of parameters in different redshifts ranges. For this reason, it is more instructive to explore the retrievability of \mlr, \fsfh, \ssfr, \aboh, \tauv\ and $\mu$ for galaxies in different redshift ranges, for example, $0.2<\zobs<0.4$ and $0.6<\zobs<0.8$. We redraw 10,000 spectral energy distributions in each of these redshift ranges from the new galaxy spectral library, which we use to generate pseudo-observations in the $ugriz$ photometric bands, for $\sn=30$. As before, we assume that the redshift \zobs\ of a galaxy is unknown a priori and use all models at redshifts between 0 and 1 in the spectral library to retrieve likelihood distributions of physical parameters. According to Fig.~\ref{fig:phothisto}b, \zobs\ can be retrieved with an uncertainty of roughly 0.05 in both redshift ranges at the adopted signal-to-noise ratio.

Fig.~\ref{fig:photzzz} shows the average probability density functions of physical parameters retrieved in this way for galaxies at redshifts $0.2<\zobs<0.4$ (top row) and $0.6<\zobs<0.8$ (bottom row). At low redshift, the constraints are similar to those obtained in Fig.~\ref{fig:photbase} for galaxies at $z=0$. The main difference is the extension of the \mlr\ probability density functions to larger values, induced by the tail of high-redshift galaxies with large K corrections in the prior distribution (Fig.~\ref{fig:disthighz}a). This tends to worsen the typical uncertainty in \mlr\ (from 0.19 to 0.22\,dex) over the range sampled by Fig.~\ref{fig:photbase}, i.e. for $\log[(\mlr)/(\mlrsun)]<0.4$, while at larger \mlr, the smaller number of models implies a significant bias in the retrieved likelihood distribution. Another noticeable difference caused by the absence of redshift information between Fig.~\ref{fig:photzzz} (top row) and Fig.~\ref{fig:photbase} is the slightly worse uncertainty and bias in the retrieved specific star formation rate of actively star-forming galaxies. Fig.~\ref{fig:photzzz} (bottom row) also shows that, at higher redshift, the loss of \ha\ information further worsens \ssfr\ estimates in these galaxies. The constraints on \mlr\ are similar to those for low-redshift galaxies. In both redshift ranges in Fig.~\ref{fig:photzzz}, the parameters \fsfh, \aboh, \tauv\ and $\mu$ are as poorly constrained as for $z=0$ galaxies in Fig.~\ref{fig:photbase}. 

We conclude that high-quality photometry in the $ugriz$ bands can provide useful constraints on the mass-to-light ratio and specific star formation rate of galaxies, especially at low redshift, but less so on the recent history of star formation and the enrichment in metals and dust of the interstellar medium. The lack of redshift information can significantly weaken these constraints.

\subsection{Spectroscopic constraints}
\label{sec:spec}

We now investigate how the constraints obtained above from multi-band photometry on the parameters \mlr, \fsfh, \ssfr, \aboh, \tauv\ and $\mu$ can be improved by appealing to different types of spectroscopic observations. In Section~\ref{sec:ew} below, we first show that the equivalent widths of strong optical emission lines alone can provide useful clues about the specific star formation rate and enrichment of the interstellar medium in star-forming galaxies. Then, in Sections~\ref{sec:lowres} and \ref{sec:medres}, we explore in detail the constraints retrievable from low- and medium-resolution spectroscopy, exploiting the ability with our approach to interpret simultaneously the stellar and nebular emission from galaxies. Throughout this section, we assume that the galaxies have well-determined spectroscopic redshifts and perform all calculations at rest-frame wavelengths in the range between 3600 and 7400\,\AA\ (Fig.~\ref{fig:methods}).

\subsubsection{Equivalent widths of strong optical emission lines}
\label{sec:ew}

The equivalent widths of optical emission lines reflect the ratio of young stars, which produce ionizing radiation, to older stars, which dominate the optical continuum radiation. These equivalent widths should therefore provide valuable information about the specific star formation rate and, if several lines are available, the gas-phase oxygen abundance and dust attenuation optical depth in star-forming galaxies. Since line equivalent widths do not provide any information about the absolute stellar continuum emission, they cannot be used to constrain the parameters \mlr\ nor \fsfh. Similarly, line equivalent widths should not provide any useful constraint on the dust attenuation optical depth in the ambient ISM, $\mu\tauv$, which affects in similar ways emission lines emerging from stellar birth clouds and the continuum radiation from old stars (equation~\ref{eq:taulism}). We therefore focus here on the retrievability of the parameters \ssfr, \aboh\ and the dust attenuation optical depth in stellar birth clouds, $(1-\mu)\tauv$, from observations of the equivalent widths of strong optical emission lines in star-forming galaxies.

To proceed, we select a sample of 10,000 spectral energy distributions from the library assembled in Section~\ref{sec:cats} to compute pseudo-observations of the net emission equivalent widths of the following emission lines: $\oii\lambda3727$; \hb; $\oiii\lambda4959$; $\oiii\lambda5007$; $\nii\lambda6548$; \ha; $\nii\lambda6584$; $\sii\lambda6716$ and $\sii\lambda6731$. We consider both the cases of low (50\,\AA\  FWHM; $R=100$ at $\lambda=5000\,$\AA) and medium (5\,\AA\  FWHM; $R=1000$ at $\lambda=5000\,$\AA) spectral resolution and fix the median signal-to-noise ratio per pixel at $\snmed=20$.  In the case of low spectral resolution, we treat as single emission features the line blends: $\oiii\lambda\lambda4959,5007$; $\nii\lambda6548+\ha+\nii\lambda6584$ and $\sii\lambda\lambda6716,6731$. We also require that, at a given spectral resolution, the net emission equivalent widths of all features be measured at better than the 3$\sigma$ level.\footnote{\label{foot:eqwidth}To compute net emission equivalent widths in low-resolution (medium-resolution) spectra, we measure the continuum level as follows. We first de-spike the spectrum with a 9-pixel (11-pixel) median smoothing. We then mask 5-pixel-wide (9-pixel-wide) regions  centred on the emission lines, through which we interpolate the spectrum. Finally, we smooth the spectrum again and adopt the result as the continuum level.} We apply the same requirement to the model spectra used to compute likelihood distributions of \ssfr, \aboh\ and $(1-\mu)\tauv$ for each pseudo-galaxy. These requirements are filled by roughly 2.7 million out of the 5 million original models at low resolution, and 1.7 million models at medium resolution (where lines are not blended). 

\begin{figure}
\begin{center}
\includegraphics[width=0.53\textwidth]{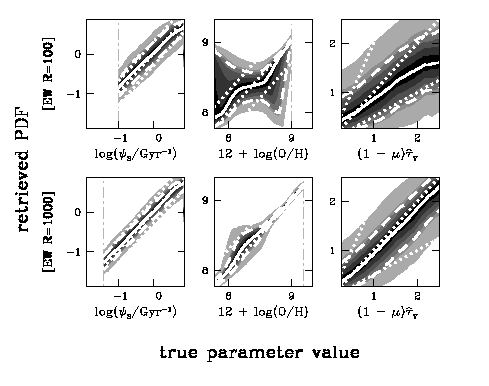}
\caption{Average probability density functions of the specific star formation rate, \ssfr, gas-phase oxygen abundance, \aboh, and dust attenuation optical depth in stellar birth clouds,  $(1-\mu)\tauv$, retrieved, using the Bayesian approach described in Section~\ref{sec:stat}, from the equivalent widths of optical emission lines in a sample of 10,000 star-forming pseudo-galaxies. The sample was extracted from the library of galaxy spectral energy distributions assembled in Section~\ref{sec:cats}, assuming a median signal-to-noise ratio per pixel  $\snmed=20$ and requiring 3$\sigma$ measurements of all line equivalent widths. ({\it Top row}) using the equivalent widths of $\oii\lambda3727$; \hb; $\oiii\lambda\lambda4959,5007$; $\nii\lambda6548+\ha+\nii\lambda6584$; and $\sii\lambda\lambda6716,6731$ at a spectral resolution of 50\,\AA\  FWHM ($R=100$ at $\lambda=5000\,$\AA). ({\it Bottom row})  using the equivalent widths of $\oii\lambda3727$; \hb; $\oiii\lambda4959$; $\oiii\lambda5007$; $\nii\lambda6548$; \ha; $\nii\lambda6584$; $\sii\lambda6716$ and $\sii\lambda6731$ at a spectral resolution of 5\,\AA\  FWHM ($R=1000$ at $\lambda=5000\,$\AA). In all panels, the lines and shading have the same meaning as in Fig.~\ref{fig:photbase}.}
\label{fig:ew}
\end{center}
\end{figure}

Fig.~\ref{fig:ew} shows the average retrieved probability density functions of \ssfr, \aboh\ and $(1-\mu)\tauv$ for the 10,000 star-forming pseudo-galaxies in our sample, for low (top row) and medium (bottom row) spectral resolution. We find that, as expected, the equivalent widths of strong optical emission lines set tight constraints on the specific star formation rate, \ssfr. The uncertainty in the retrieved best estimate is only slightly worse at low (0.23\,dex) than at medium (0.18\,dex) spectral resolution. An advantage of medium spectral resolution is the possibility to constrain \ssfr\ in more quiescent star-forming galaxies than at low spectral resolution, for fixed \snmed. According to Fig.~\ref{fig:ew}, for $\snmed=20$, the limit under which the equivalent widths of all lines of interest cannot be measured at the 3$\sigma$ level is $\ssfr\approx0.1\,\textrm{Gyr}^{-1}$ at $R=100$ and $\ssfr\approx0.04\,\textrm{Gyr}^{-1}$ at $R=1000$.

Fig.~\ref{fig:ew} (bottom row) further shows that, at medium spectral resolution, the equivalent widths of optical emission lines  provide tight constraints on the gas-phase oxygen abundance in star-forming galaxies. The uncertainty in the retrieved best estimate of \aboh\ is only 0.02 at the high-metallicity end. It reaches 0.17 at low \aboh, because the rise in intrinsic emission-line intensity with decreasing metallicity \citep[e.g., figure~4d of][]{charlot2001} allows the inclusion of more dusty galaxies in the likelihood distributions. At low spectral resolution, the blending of the \ha\ and \nii\ lines worsens considerably these constraints (Fig.~\ref{fig:ew}, top row). Moreover, the drop in emission-line intensity with increasing metallicity implies that \aboh\ can be retrieved in more metal-rich galaxies at medium than at low spectral resolution, for fixed \snmed. The constraints on the dust attenuation optical depth in stellar birth clouds in Fig.~\ref{fig:ew} are also much tighter at medium than at low spectral resolution. At low spectral resolution, the median retrieved uncertainty in  $(1-\mu)\tauv$ is only about 20 percent larger than at medium spectral resolution, but the best estimate is much more severely biased toward the prior expectation.

\subsubsection{Low-resolution spectroscopy}
\label{sec:lowres}

We now turn to one of the most innovative parts of the present study enabled by the approach developed in Section~\ref{sec:models}: the retrieval of physical parameters from the simultaneous interpretation of strong absorption and emission features in low-resolution spectral energy distribution of galaxies over the entire rest-wavelength range from $\lambda=3600$ to 7400\,\AA. As before, we extract 10,000 spectral energy distributions from the library assembled in Section~\ref{sec:cats} to compute pseudo-observations of galaxies at the resolution of 50\,\AA\  FWHM ($R=100$ at $\lambda=5000\,$\AA) in this wavelength range. We adopt for the moment a median signal-to-noise ratio per pixel $\snmed=20$. We use the rest of the 5 million models in the library to retrieve likelihood distributions of physical parameters for each of these 10,000 pseudo-galaxies.

\begin{figure*}
\begin{center}
\includegraphics[width=\textwidth]{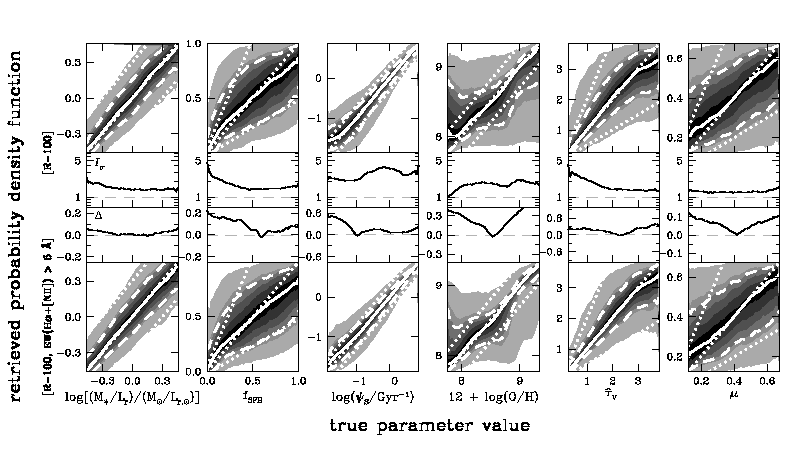}
\caption{Average probability density functions of the same 6 physical parameters as in Fig.~\ref{fig:dist} retrieved, using the Bayesian approach described in Section~\ref{sec:stat}, from low-resolution optical spectra of a sample of 10,000 pseudo-galaxies. The spectra cover the  wavelength range from $\lambda=3600$ to 7400\,\AA\ at the resolution of 50\,\AA\  FWHM ($R=100$ at $\lambda=5000\,$\AA) with median signal-to-noise ratio per pixel $\snmed=20$. ({\it Top row}) for a sample of 10,000 pseudo-galaxies extracted randomly from the spectral library assembled in Section~\ref{sec:cats}. The improvement factor \isig\ and the gain in accuracy $\Delta$ (equations~\ref{eq:isig}--\ref{eq:delta}) are shown as a function of true parameter value at the bottom of each panel to quantify differences in the retrieved likelihood distributions relative to the standard case of Fig.~\ref{fig:photbase}. ({\it Bottom row}) for a sample of 10,000 star-forming pseudo-galaxies extracted from the same spectral library, with the requirement that the net \ha+\nii\ emission equivalent width be greater than 5\,\AA. In all panels, the lines and shading have the same meaning as in Fig.~\ref{fig:photbase}.}
\label{fig:specr100}
\end{center}
\end{figure*}

Fig.~\ref{fig:specr100} (top row) shows the average retrieved probability density functions of \mlr, \fsfh, \ssfr, \aboh, \tauv\ and $\mu$ for the 10,000 pseudo-galaxies in our sample. Also shown at the bottom of each panel are the factor of improvement in the retrieved uncertainty, \isig, and the gain in accuracy, $\Delta$, relative to the standard case of Fig.~\ref{fig:photbase} (equations~\ref{eq:isig}--\ref{eq:delta}). Fig.~\ref{fig:specr100} (top row) reveals that a major improvement of low-resolution spectroscopy over rest-frame $ugriz$ photometry is the ability to retrieve meaningful (i.e. unbiased) constraints not only on the stellar mass-to-light ratio and the specific star formation rate, but also on the recent star formation history (\fsfh) and the enrichment of metals and dust in the interstellar medium of galaxies [\aboh, \tauv\ and, to a lesser extent, $\mu$]. The gain in accuracy ($\Delta$) reaches up to 30 percent of the explored dynamic range for \aboh, 15 percent for \tauv\ and $\mu$ and up to 20 percent for \fsfh. This results from the ability to detect and interpret, even at low resolution, the strong Balmer absorption features of intermediate-age stars and the emission-line signatures of interstellar gas. The median uncertainty in the retrieved fraction of current stellar mass formed in the last 2.5\,Gyr amounts to about 0.23, that in \aboh\ about 0.28 and that in the total dust attenuation optical depth, \tauv, about 0.64. For \mlr\ and \ssfr, already constrained in a meaningful way by multiband photometry, the improvement in uncertainty relative to the standard case amounts to a factor (\isig) of between 1.5 and 3.

A more in-depth investigation reveals that, in all panels in Fig.~\ref{fig:specr100} (top row), the largest uncertainties pertain to galaxies with emission lines too weak to allow a distinction between the signatures of star formation history, metallicity and dust in the spectra. These are mainly galaxies with low specific star formation rate, high metallicity and strong attenuation of emission lines by dust in stellar birth clouds. In this context, it is interesting to explore how much the constraints tighten if we restrict the sample to low-resolution galaxy spectra exhibiting unambiguous emission lines. We appeal again to the library assembled in Section~\ref{sec:cats} to generate a sample of 10,000 pseudo-observations of low-resolution galaxy spectra, for $\snmed=20$, with the requirement that the net \ha+\nii\ emission equivalent width be greater than 5\,\AA\ (throughout this paper, we adopt the convention of positive equivalent widths for emission lines). Fig.~\ref{fig:specr100} (bottom row) shows that, in this case, the constraints on the interstellar parameters \aboh\ and \tauv\ improve significantly. The corresponding median uncertainties both drop by about 20 percent. The constraints on \ssfr\ improve mainly at the low end of the explored range (the uncertainty dropping by about 15 percent at $\ssfr\la0.1\,\textrm{Gyr}^{-1}$), where the omission of galaxies with weak emission lines is most significant. The improvement is only marginal for \mlr, \fsfh\ and $\mu$, which are constrained primarily by stellar continuum and absorption-line features rather than by nebular emission lines.

\begin{figure*}
\begin{center}
\includegraphics[width=\textwidth]{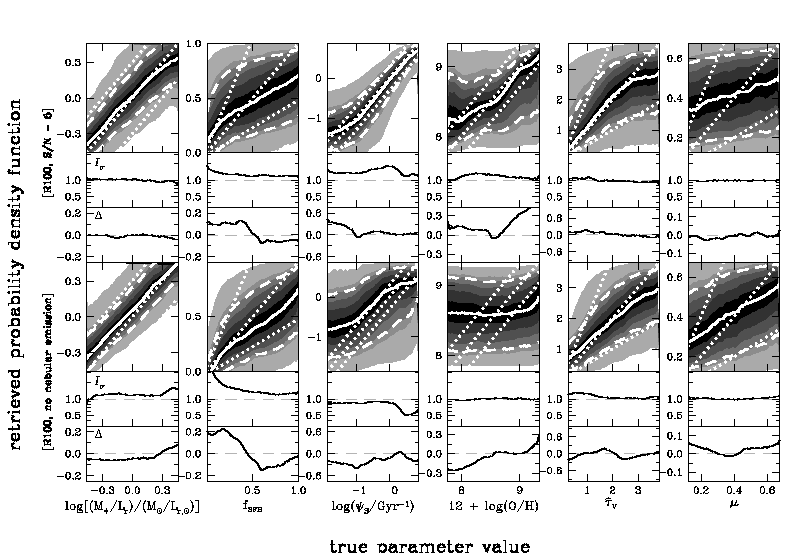}
\caption{Same as Fig.~\ref{fig:specr100}, but: ({\it top row}) adopting a median signal-to-noise ratio per pixel $\snmed=5$ instead of 20;  ({\it bottom row}) not including nebular emission in the model library used to analyze the sample of 10,000 pseudo-galaxies. In both cases, the improvement factor \isig\ and the gain in accuracy $\Delta$ (equations~\ref{eq:isig}--\ref{eq:delta}) are shown as a function of true parameter value at the bottom of each panel to quantify differences in the retrieved likelihood distributions relative to the standard case of Fig.~\ref{fig:photbase}.}
\label{fig:specr100deg}
\end{center}
\end{figure*}

The ability with our approach to exploit strong absorption and emission features in low-resolution galaxy spectra depends critically on the quality of the observations. To illustrate this, we show in Fig.~\ref{fig:specr100deg} (top row) the analog of Fig.~\ref{fig:specr100} (top row) when adopting a median signal-to-noise ratio per pixel $\snmed=5$ instead of 20. The reduced ability to interpret spectral features at $\snmed=5$ makes the median uncertainties in the different retrieved parameters increase by 30 percent (for \mlr, \fsfh, $\mu$) up to 90 percent (for \ssfr) relative to the case $\snmed=20$. It also introduces substantial biases in the retrieved likelihood distributions. A comparison of Fig.~\ref{fig:specr100deg} (top row) with Fig.~\ref{fig:photbase} reveals that, in fact, low-resolution spectroscopy with $\snmed=5$ over the wavelength range from $\lambda=3600$ to 7400\,\AA\ provides hardly better constraints on the different parameters than high-quality $ugriz$ photometry (as confirmed by the fact that \isig\ is close to unity and $\Delta$ close to zero in Fig.~\ref{fig:specr100}, top row). The constraints on \ssfr\  and \aboh\ are somewhat tighter because of the influence of actively star-forming galaxies with strong emission lines on the average retrieved probability density functions. The star formation history parameter is only slightly better constrained, as even strong Balmer absorption lines are difficult to interpret at low \snmed. The constraints on \tauv\ are as poor as inferred from multiband photometry, reflecting the inability to extract useful information from the \ha/\hb\ ratio in low-quality spectra. In Fig.~\ref{fig:specr100deg} (top row), the constraints on \mlr\ are actually weaker and more severely biased at the extremities of the explored range than those obtained from rest-frame $ugriz$ photometry in Fig.~\ref{fig:photbase}. This suggests that high-quality photometry including the near-infrared $iz$ bands can better constrain galaxy stellar masses than low-resolution optical spectroscopy with $\snmed=5$.

The inclusion of nebular emission also has a crucial influence on the ability to constrain galaxy physical parameters from low-resolution optical spectroscopy. Fig.~\ref{fig:specr100deg} (bottom row) shows the results obtained when attempting to interpret the same pseudo-observations as in Fig.~\ref{fig:specr100} (top row; which include nebular emission) with the library of models not including nebular emission already used in Fig.~\ref{fig:photdeg} (bottom row).\footnote{\label{foot:despike} To avoid artefacts in the resulting probability density functions, we remove prominent emission lines from the pseudo-galaxy spectra before computing the likelihood of models without nebular emission (equation~\ref{eq:prob}). To achieve this, we first measure the continuum level across the whole range from $\lambda=3600$ to 7400\,\AA, as described in footnote~\ref{foot:eqwidth}. Then, we reset to the continuum level the flux of every pixel exceeding the continuum flux by more than 5 times the noise level.} In this case, the average retrieved likelihood distributions of all parameters are severely degraded relative to Fig.~\ref{fig:specr100} (top row). In fact, for most parameters, the constraints are worse than those derived in Fig.~\ref{fig:photbase} from high-quality $ugriz$ photometry when accounting for the effect of nebular emission on photometric fluxes. Fig.~\ref{fig:specr100deg} (bottom row) shows that, for example, using models which neglect the contamination of the $r$-band flux by \ha+\nii\ and \sii\ line emission biases \mlr\ estimates upward for star-forming galaxies. Moreover, ignoring the contamination of stellar Balmer absorption features by nebular emission in low-resolution spectra causes systematic underestimates of \fsfh\ in star-forming galaxies. For actively star-forming galaxies, the constraints on \ssfr\ in Fig.~\ref{fig:specr100deg} (bottom row) are weaker than the rough constraints derived in Fig.~\ref{fig:photbase} from the influence of nebular emission on $ugriz$ photometric observations. Not surprisingly, models without nebular emission do not allow the extraction of useful constraints on the interstellar parameters \aboh, \tauv\ and $\mu$ from low-resolution galaxy spectra.

Hence, we find that important constraints can be obtained on the stellar and interstellar parameters of galaxies from low-resolution spectra at optical wavelengths. This however requires high-quality observations as well as the inclusion of nebular emission in the spectral analysis.

\begin{figure*}
\begin{center}
\includegraphics[width=\textwidth]{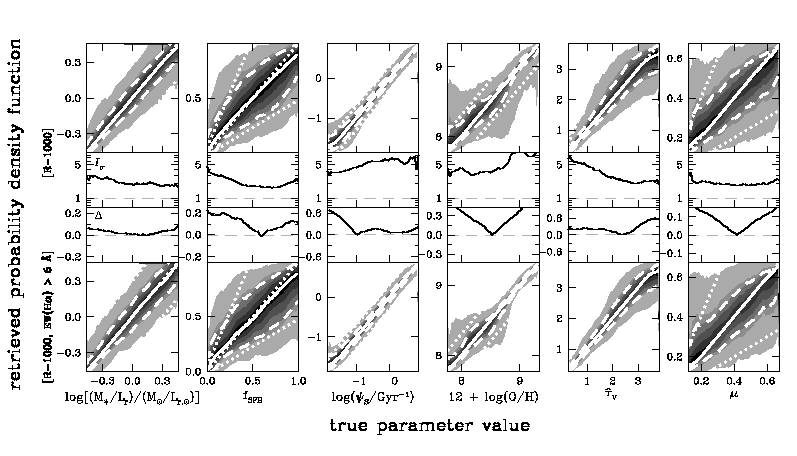}
\caption{Average probability density functions of the same 6 physical parameters as in Fig.~\ref{fig:dist} retrieved, using the Bayesian approach described in Section~\ref{sec:stat}, from medium-resolution optical spectra of a sample of 10,000 pseudo-galaxies. The spectra cover the  wavelength range from $\lambda=3600$ to 7400\,\AA\ at the resolution of 5\,\AA\  FWHM ($R=1000$ at $\lambda=5000\,$\AA) with median signal-to-noise ratio per pixel $\snmed=20$. ({\it Top row}) for a sample of 10,000 pseudo-galaxies extracted randomly from the spectral library assembled in Section~\ref{sec:cats}. The improvement factor \isig\ and the gain in accuracy $\Delta$ (equations~\ref{eq:isig}--\ref{eq:delta}) are shown as a function of true parameter value at the bottom of each panel to quantify differences in the retrieved likelihood distributions relative to the standard case of Fig.~\ref{fig:photbase}. ({\it Bottom row}) for a sample of 10,000 star-forming pseudo-galaxies extracted from the same spectral library, with the requirement that the net \ha\ emission equivalent width be greater than 5\,\AA. In all panels, the lines and shading have the same meaning as in Fig.~\ref{fig:photbase}.}
\label{fig:specr1000}
\end{center}
\end{figure*}

\subsubsection{Medium-resolution spectroscopy}
\label{sec:medres}

Medium-resolution spectroscopy should provide even more insight into the physical properties of galaxies than low-resolution spectroscopy, because composite features such as the \oiii\ doublet, the \ha+\nii\ blend and the \sii\ doublet can be resolved into individual emission lines. To investigate this, we extract 10,000 spectral energy distributions from the library assembled in Section~\ref{sec:cats} and compute pseudo-observations of galaxies at the resolution of 5\,\AA\  FWHM ($R=1000$ at $\lambda=5000\,$\AA) over the wavelength range from $\lambda=3600$ to 7400\,\AA. We adopt a median signal-to-noise ratio per pixel $\snmed=20$. As before, we use the rest of the 5 million models in the library to retrieve likelihood distributions of physical parameters for each of these 10,000 pseudo-galaxies.

In Fig.~\ref{fig:specr1000} (top row), we show the average probability density functions of \mlr, \fsfh, \ssfr, \aboh, \tauv\ and $\mu$ retrieved from the medium-resolution spectra of the 10,000 pseudo-galaxies in our sample. The factor \isig\ and the gain $\Delta$ quantifying the improvements  in the retrieved uncertainty and accuracy relative to the standard case of Fig.~\ref{fig:photbase} (equations~\ref{eq:isig}--\ref{eq:delta}) are also shown at the bottom of each panel. A comparison of Fig.~\ref{fig:specr1000} (top row) with Fig.~\ref{fig:specr100} (top row) shows that an increase in spectral resolution by a factor of 10 produces significantly tighter constraints on all parameters. The improvement is most significant for the specific star formation rate, the gas-phase oxygen abundance and the total dust attenuation optical depth, for which the median retrieved uncertainty drops by about 45 percent from low to medium spectral resolution. The gain is more modest, of the order of 25 percent, for \mlr, \fsfh, and $\mu$. The refined spectral information available at medium resolution also help reduce some biases in the retrieved best estimates of \fsfh\ (for young galaxies), \ssfr\ (for quiescent star-forming galaxies) and $\mu$ (across most of the explored range) in Fig.~\ref{fig:specr1000} (top row) relative to Fig.~\ref{fig:specr100} (top row).  

By analogy with Fig.~\ref{fig:specr100}, we further show in Fig.~\ref{fig:specr1000} (bottom row) the average probability density functions obtained when selecting a sample of 10,000 pseudo-observations of medium-resolution galaxy spectra, for $\snmed=20$, with the requirement that the net \ha\ emission equivalent width be greater than 5\,\AA. The retrieved likelihood distributions of \aboh\ and \tauv\ are significantly narrower than when including also galaxies with weak emission lines in the sample, with median retrieved uncertainties of about 0.08 for \aboh\ and 0.30 for \tauv. As at low spectral resolution, restricting the sample to galaxies with prominent emission lines reduces the uncertainties in \ssfr\ mainly at the low end of the explored range. Changes in the retrieved likelihood distributions of \mlr, \fsfh\ and $\mu$ are marginal, because these parameters are constrained primarily by stellar continuum and absorption-line features.

We have also investigated how the constraints obtained in Fig.~\ref{fig:specr1000} from the analysis of medium-resolution galaxy spectra depend on the assumed signal-to-noise ratio of the observations and the inclusion of nebular emission in the modeling (not shown). We find that adopting a median signal-to-noise ratio per pixel $\snmed=5$ instead of 20 for the 10,000 pseudo-galaxy spectra makes the median retrieved uncertainties in most parameters increase by 30 to 50 percent (only 15 percent for \mlr) relative to Fig.~\ref{fig:specr1000} (top row). The biases in the retrieved best estimates are less severe than found in Fig.~\ref{fig:specr100deg} (top row) at low spectral resolution. As expected from  Fig.~\ref{fig:specr100deg} (bottom row), accounting for nebular emission also has a critical influence on the retrievability of the specific star formation rate and the interstellar parameters from medium-resolution galaxy spectra. It is worth pointing out that, at medium (and high) spectral resolution, prominent nebular emission lines can be excised from the spectra and analyzed separately from the stellar continuum and absorption-line signatures if no tool is available to analyze the stellar and nebular emission simultaneously \citep[e.g.,][]{kauffmann2003a,brinchmann2004,gallazzi2005}. This is not the case at low spectral resolution, where the analysis of the stellar and nebular emission must be combined to obtain meaningful constraints on galaxy physical parameters.


\section{Application to a real sample}
\label{sec:sample}

The conclusions drawn in the previous section about the retrievability of galaxy physical parameters from different types of observations at optical wavelengths rely on sophisticated simulations of galaxy spectral energy distributions. As emphasized in Section~\ref{sec:stat}, such pseudo-observations are required to test the retrievability of physical parameters from observables, because the physical parameters of real  galaxies cannot be known a priori. In this section, we apply our approach to the interpretation of observed medium-resolution spectra of galaxies from the SDSS DR7. We compare the results obtained through the simultaneous interpretation of stellar and nebular emission (at the resolving power $R\approx1000$ considered in Section~\ref{sec:medres}) with results obtained for the same galaxies from previous separate analyses of nebular emission-line and stellar absorption-line features (at the native resolving power $R\approx2000$ of the SDSS observations). We also investigate the influence of the adopted prior distributions on the retrieved physical parameters of the galaxies.

\subsection{Physical parameters of SDSS galaxies}
\label{sec:sdss}

The SDSS DR7 contains about 1 million galaxy spectra covering the wavelength range from 3800 to 9200\,{\AA} (observer frame) at a spectral resolution of roughly 3\,\AA\ FWHM. Of these, we extract a subsample of 12,660 star-forming galaxies with high-quality spectra as follows. We first reject any duplicate spectrum of a same galaxy, all spectra of merger candidates (i.e. corresponding to spatial separations less than 1.5\arcsec\ and redshift separations less than 200\,km\,s$^{-1}$) and all galaxies at redshift less than $z=0.03$ (to ensure that the spectrum includes the $\oii\lambda3727$ doublet) from the original sample. Among the remaining 829,570 galaxies, we select all emission-line galaxies with signal-to-noise ratio greater than 10 in $\oii\lambda3727$, \hb, $\oiii\lambda5007$, \ha, $\nii\lambda6584$ and $\sii\lambda6716$. We reject AGNs by applying the conservative criterion of \citet{kauffmann2003b} in the standard \citet{baldwin1981} line-diagnostic diagram (see Fig.~\ref{fig:lineratios}a above). The resulting 28,075 star-forming galaxies have a mean redshift of $0.08\pm0.05$. We exclude all galaxies at redshift $z>0.15$ to keep the sample homogeneous and further select galaxies with median signal-to-noise ratio per pixel in the range $5<\snmed<25$. This leaves 23,213 spectra of galaxies at a mean redshift of $0.07\pm0.03$, all of which have net \ha\ emission equivalent widths greater than 5\,\AA. We identify 12,660 galaxies in this sample for which good measurements of stellar mass, star formation rate, gas-phase oxygen abundance and dust attenuation optical depth are available from previous studies (see next paragraph). For the purpose of illustrating the usefulness of our approach to interpret galaxy spectra, we focus on the galaxy physical properties probed by the 3-arcsec-diameter SDSS fibers and do not consider any photometry-based aperture correction.
 
Estimates of star formation rate, gas-phase oxygen abundance and dust attenuation optical depth are available for the 12,600 SDSS galaxies in this sample from the work of \citet{brinchmann2004} and \citet{tremonti2004}. These estimates were obtained by fitting the fluxes of the $\oii\lambda3727$, \hb, $\oiii\lambda5007$, \ha, $\nii\lambda6584$ and $\sii\lambda6716$ emission lines (measured to high precision by adjusting a stellar population model to the spectral continuum) with a large library of  \citet{charlot2001} nebular emission models. In addition, stellar-mass estimates are available for all galaxies from the work of \citet{gallazzi2005}, who fitted the strengths of the 4000\,\AA\ break and the \hb, \hdg, \mgfep\ and \mgtfe\ absorption-line features with a large library of \citet{bruzual2003} stellar population synthesis models. The combination of these stellar-mass estimates with the star-formation-rate estimates of \citet{brinchmann2004} provides estimates of the specific star formation rate probed by the SDSS fiber. 

The above physical parameters were derived from spectral analyses at the original SDSS resolution of about 3\,\AA\ FWHM. To exemplify the power of the approach developed in Section~\ref{sec:models}, and for the purpose of comparison with the results of Section~\ref{sec:medres}, we degrade the SDSS spectra to a resolution of 5\,\AA\  FWHM, adopting as before a fixed pixel size of 2.5\,\AA. Also, we focus on the rest-frame wavelength range from 3700 to 7400\,\AA, which does not extend as much in the ultraviolet as that considered in Section~\ref{sec:medres} (Fig.~\ref{fig:methods}a), but which does include the $\oii\lambda3727$ emission doublet. By analogy with the analysis of pseudo-observations in Section~\ref{sec:medres}, we use the 5 million models in the spectral library assembled in Section~\ref{sec:cats} to retrieve likelihood distributions of physical parameters from the simultaneous analysis of the stellar and nebular emission (equations~\ref{eq:prob}--\ref{eq:relprob}) for each SDSS galaxy in the sample. We note that, when computing likelihood distributions for a given galaxy, we exclude from the library all models older than the age of the Universe at the redshift of that galaxy.

\begin{figure*}
\begin{center}
\includegraphics[width=0.35\textwidth]{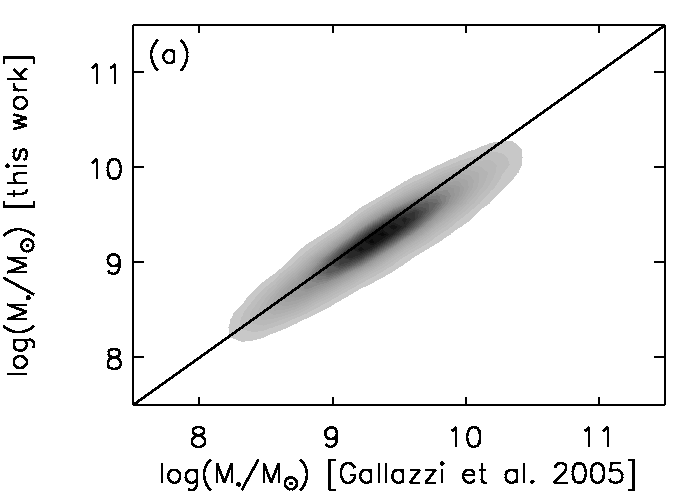}
\includegraphics[width=0.35\textwidth]{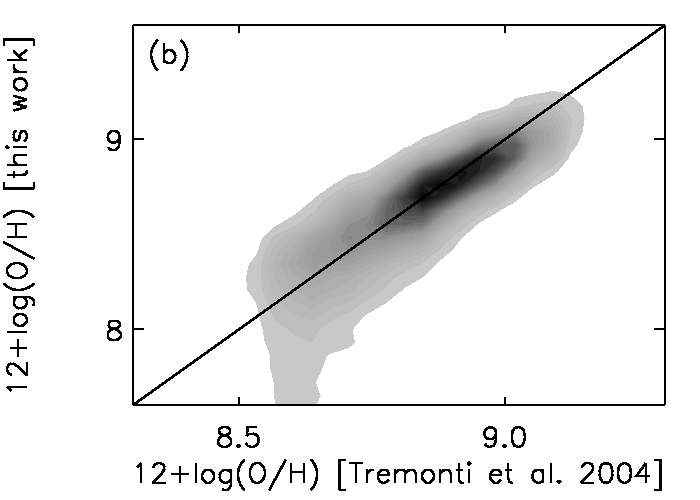}
\includegraphics[width=0.35\textwidth]{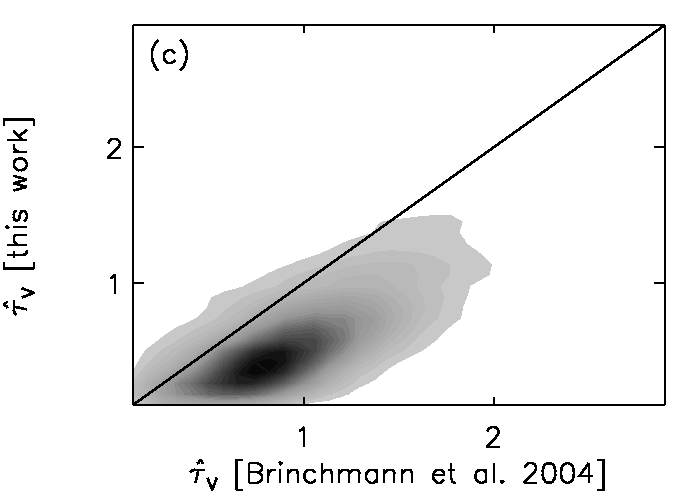}
\includegraphics[width=0.35\textwidth]{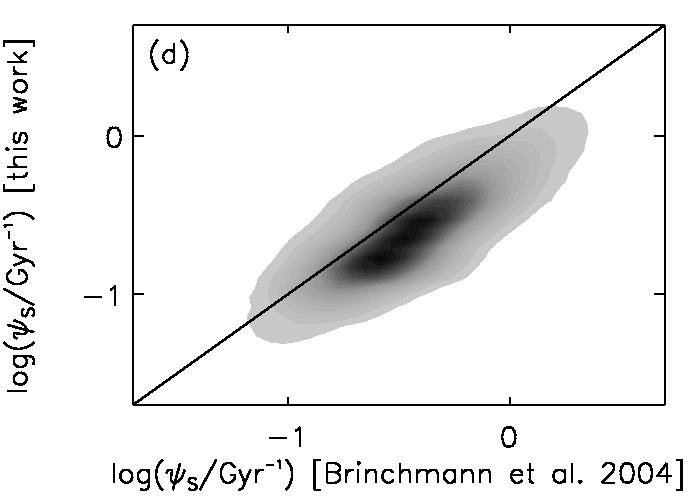}
\caption{Estimates of physical parameters retrieved from the spectra of 12,660 SDSS star-forming galaxies (degraded to a resolution of 5\,\AA\  FWHM) using the Bayesian approach described in Section~\ref{sec:stat} plotted against estimates of the same parameters from different sources, as indicated. ({\it a}) Stellar mass, \mstar. ({\it b}) Gas-phase oxygen abundance, \aboh. ({\it c}) Total $V$-band attenuation optical depth of the dust, \tauv.  ({\it d}) Specific star formation rate, \ssfr. In each panel, the likelihood distributions from previous studies were reconstructed from the published 16th, 50th and 84th percentiles and combined with the likelihood distributions obtained in this work to generate 2D probability density functions. The contours depict the normalized co-added 2D probability density function including all 12,660 galaxies in the sample (on a linear scale 25 levels, the outer edge of the lightest grey level corresponding to the 96th percentile of the 2D distribution). The solid line is the identity relation.}
\label{fig:r1000jar}
\end{center}
\end{figure*}

In Fig.~\ref{fig:r1000jar}, we show how the probability density functions of stellar mass, gas-phase oxygen abundance, total $V$-band attenuation optical depth of the dust and specific star formation rate retrieved from our analysis compare with those obtained in previous separate studies of the stellar and nebular emission at higher spectral resolution. Fig.~\ref{fig:r1000jar}a shows that, for example, our estimates of \mstar\ (obtained by multiplying \mlr\ by the absolute galaxy luminosity in the observed $r$ band) agree well with those derived by \citet{gallazzi2005} from the analysis of selected absorption-line indices. The slightly lower values derived here ($\sim0.1\,$dex) are attributable to the updated stellar population synthesis prescription (Section ~\ref{sec:stars}). We also find good agreement in Fig.~\ref{fig:r1000jar}b  between our estimates of \aboh\ and those obtained by \citet{tremonti2004} from the analysis of selected emission-lines luminosities. The discrepancies at $\aboh<8.7$ are consistent with the uncertainty expected at low metallicity from interpretations of medium-resolution galaxy spectra (Fig.~\ref{fig:specr1000}, bottom row). In Figs~\ref{fig:r1000jar}c and \ref{fig:r1000jar}d, the likelihood distributions retrieved from our analysis favor systematically lower \tauv\ and \ssfr\ than derived by \citet{brinchmann2004} from the analysis of emission-line luminosities. This is because \cite{brinchmann2004} adopt a shallower dust attenuation curve ($\taul\propto\lambda^{-0.7}$ both in stellar birth clouds and in the ambient ISM) than that resulting typically from equations~(\ref{eq:taulbc})--(\ref{eq:slope}). This makes the inferred dust optical depth and the correction for obscured star formation smaller at fixed observed reddening.

Fig.~\ref{fig:r1000jar} therefore shows that the likelihood distributions of \mstar, \aboh, \tauv\ and \ssfr\ retrieved using our approach agree well (modulo some understood offsets) with those obtained in previous separate analyses of the stellar and nebular emission of SDSS galaxies at twice higher spectral resolution. Such an agreement is expected in part from the fact that these previous analyses rely on earlier versions of the same stellar population synthesis code and nebular emission models as adopted here (\citealt{bruzual2003,charlot2001}; see Sections~\ref{sec:stars} and \ref{sec:neb}). What is most remarkable is the demonstrated ability with our approach to interpret simultaneously the stellar and nebular emission from galaxies without having to excise nebular emission lines from noisy spectra. As shown in Section~\ref{sec:estim}, this ability extends to low-resolution galaxy spectra and even to multi-band photometry.

\begin{figure*}
\begin{center}
\includegraphics[width=0.177\textwidth]{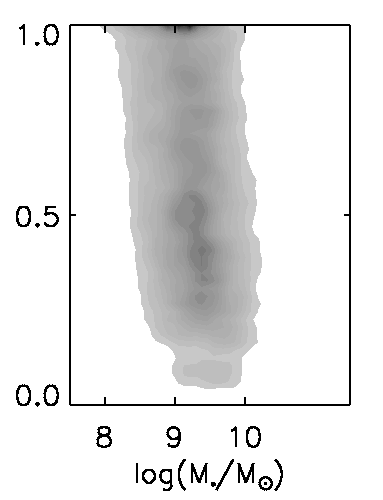}
\includegraphics[width=0.177\textwidth]{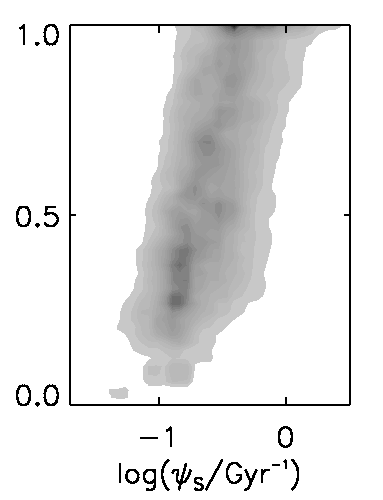}
\includegraphics[width=0.177\textwidth]{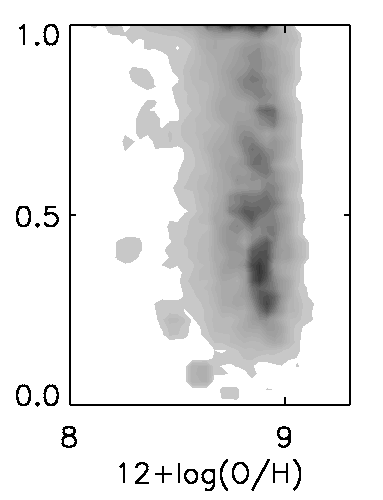}
\includegraphics[width=0.177\textwidth]{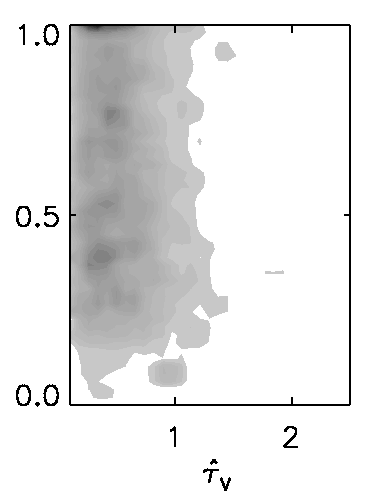}
\includegraphics[width=0.177\textwidth]{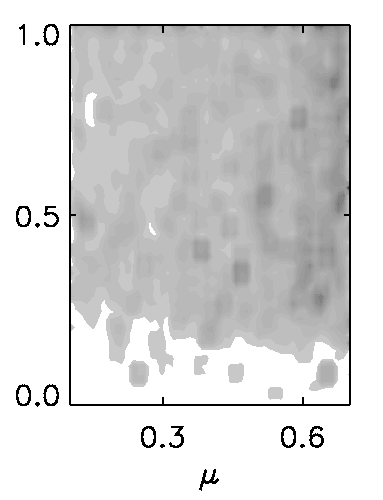}
\caption{Two-dimensional probability density functions of the fraction of the current stellar mass formed during the last 2.5\,Gyr, \fsfh, versus stellar mass, $\mstar$, specific star formation rate, $\ssfr$, gas-phase oxygen abundance, \aboh, total $V$-band attenuation optical depth of the dust, \tauv, and fraction of \tauv\ arising from dust in the ambient ISM, $\mu$. All parameters were retrieved from the spectra of the same 12,660 SDSS star-forming galaxies as in Fig.~\ref{fig:r1000jar} (degraded to a resolution of 5\,\AA\  FWHM) using the Bayesian approach described in Section~\ref{sec:stat}. In each panel, the contours depict the normalized co-added 2D probability density function including all 12,660 galaxies in the sample (on a linear scale of 25 levels, the outer edge of the lightest grey level corresponding to the 96th percentile of the 2D distribution). All physical quantities pertain to the regions of galaxies probed by the 3-arcsec-diameter SDSS fibers.}
\label{fig:r1000jarfsfh}
\end{center}
\end{figure*}

It is of interest to see how the fraction of the current stellar mass formed during the last 2.5\,Gyr, which has not been investigated previously, relates to the other physical parameters of SDSS galaxies retrieved using our approach. In Fig.~\ref{fig:r1000jarfsfh}, we show the 2D probability density functions of \fsfh\ versus $\log\mstar$, $\log\ssfr$, \aboh, \tauv\ and $\mu$ for the 12,660 SDSS star-forming galaxies in our sample. The Spearman rank correlation coefficients (evaluated using the median estimates of the different parameters) are $r_s=-0.27$ for the relation between \fsfh\ and $\log\mstar$ and $+0.59$ for that between \fsfh\ and $\log\ssfr$. For this sample size, both correlations are significant at better than the 5$\sigma$ level. In contrast, none of the interstellar parameters \aboh, \tauv\ and $\mu$ correlates significantly with \fsfh\ (we note in passing that the required high-\sn\ detection of 6 emission lines tends to select galaxies with low \tauv\ and high $\mu$). Taken at face value, the results of Fig.~\ref{fig:r1000jarfsfh} would suggest that the least massive galaxies, which are the most actively star-forming ones today, have also formed the largest proportion of stars over the last 2.5\,Gyr. Such a trend would be consistent with the general expectation that, as galaxies grow through star formation, the exhaustion of the gas supply makes the rate of star formation drop gradually (see also, e.g., fig.~22 of \citealt{brinchmann2004}). However, this conclusion must be taken with caution, as the present sample is not complete and the SDSS fibers probe only the inner parts of galaxies.

\subsection{Influence of the prior distributions of physical parameters}
\label{sec:prior}

The constraints derived above on the stellar and interstellar properties of SDSS galaxies, along with the analyses of pseudo-observations carried out in Section~\ref{sec:estim}, rely on the adoption of the prior distributions of galaxy physical parameters in Fig.~\ref{fig:dist}. As outlined in Section~\ref{sec:stat}, these distributions are designed to be as flat as possible to make the spectral library appropriate for the retrieval of physical parameters from observed spectral energy distributions of galaxies in wide ranges of physical properties. To illustrate the importance of this choice, we examine here the results obtained when adopting different prior distributions reflecting more specific properties of the galaxy population. As an example, we consider the prior distributions of galaxy physical parameters derived from the original semi-analytic recipes of \citet{delucia2007}.

In Section~\ref{sec:millennium}, we used the semi-analytic recipes of \citet{delucia2007} to generate the star formation and chemical enrichment histories of 500,000 galaxies extracted from the Millennium cosmological simulation of \citet{springel2005}. The parameters \mlr,  \fsfh,  \ssfr\  and the interstellar metallicity \zgas\ are computed by default in this procedure. \citet{delucia2007} do not provide any recipe to compute the gas-phase oxygen abundance nor the ionization parameter of the gas heated by young stars. Also, while attenuation by dust is included using the 2-component model of \citet[adopting a fixed attenuation curve $\taul\propto\lambda^{-0.7}$ both in stellar birth clouds and in the ambient ISM]{charlot2000}, we find that the recipe used by \citet{delucia2007} to relate the $V$-band dust optical depth to the galaxy gas mass, metallicity and orientation produces typically much lower \tauv\ than observed in SDSS galaxies. To compute prior distributions of interstellar parameters associated to the original semi-analytic recipes of \citet{delucia2007}, we therefore proceed as follows. For each of the 500,000 galaxies in the library, we draw 10 different realizations of $\log\, U_0$, \xid\ and $\mu$ from the distributions in Table~\ref{tab:params}. We also draw 10 different realizations of \tauv\ from a Gaussian distribution  centred on 1.2 with a standard deviation of 0.6. This roughly represents the global properties of SDSS galaxies for a fixed attenuation curve $\taul\propto\lambda^{-0.7}$ in stellar birth clouds and in the ambient ISM \citep{brinchmann2004}. The new spectral library contains 5 million models, which we can use to interpret observations of galaxy spectral energy distributions in the same way as the library assembled in Section~\ref{sec:cats}.

\begin{figure}
\begin{center}
\includegraphics[width=0.47\textwidth]{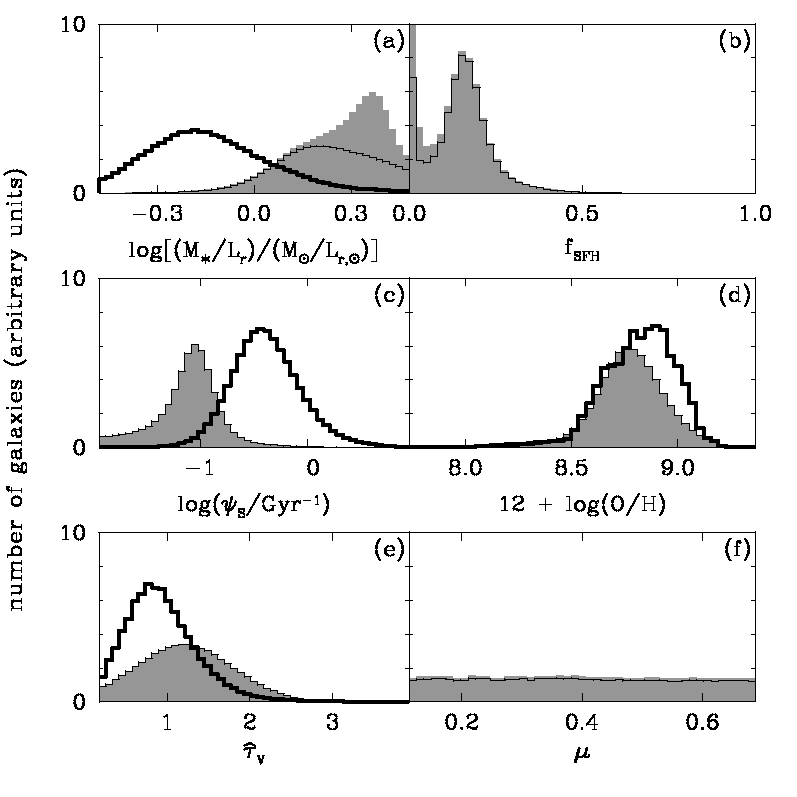}
\caption{Prior distributions of the same physical parameters as in Fig.~\ref{fig:dist} for the 5 million galaxies in the spectral library generated in Section~\ref{sec:prior}. These distributions are those arising, at the mean redshift $z=0.07$ of the SDSS sample of Section~\ref{sec:sdss},  from the original semi-analytic post-treatment of the Millennium cosmological simulation of \citet{springel2005} by \citet[see text for detail]{delucia2007}. In each panel, the shaded and thin solid histograms have the same meaning as in Fig.~\ref{fig:dist}. In panels ({\it a}), ({\it c}), ({\it d}) and ({\it e}), the thick solid histograms show the distributions of the best estimates of \mlr, \ssfr, \aboh\ and \tauv\ from \citet{gallazzi2005}, \citet{brinchmann2004} and \citet{tremonti2004} for the same sample of 12,660 SDSS star-forming galaxies as in Fig.~\ref{fig:r1000jar}.}
\label{fig:distOR}
\end{center}
\end{figure}

Fig.~\ref{fig:distOR} shows the prior distributions of \mlr, \fsfh, \ssfr, \aboh, \tauv\ and $\mu$ in this library at the mean redshift $z=0.07$ of the SDSS sample of Section~\ref{sec:sdss}. In each panel, the shaded histogram shows the distribution including all galaxies, while the thin solid histogram shows the contribution by star-forming galaxies alone. As noted in Section~\ref{sec:millennium}, 35 percent of the galaxies in the original library of \citet{delucia2007} do not form stars at low redshift, while the distributions of \ssfr\ and \fsfh\ for the remaining galaxies are narrow and characteristic of quiescent star formation (Figs~\ref{fig:distOR}b and \ref{fig:distOR}c). As a result, most galaxies have relatively large  \mlr\ (Fig.~\ref{fig:distOR}a). In Fig.~\ref{fig:distOR}d, the narrow distribution of \aboh\ arises from the correlation of the interstellar metallicity \zgas\ with \ssfr\ for star-forming galaxies in this library. Also shown as thick solid histograms in Fig.~\ref{fig:distOR} are the distributions of the best estimates of \mlr, \ssfr, \aboh\ and \tauv\ from \citet{gallazzi2005}, \citet{brinchmann2004} and \citet{tremonti2004} for the 12,660 SDSS star-forming galaxies of Section~\ref{sec:sdss}. The required detection of 6 emission lines with $\sn>10$ in these galaxies implies that they are typically more actively star-forming (and hence have lower \mlr) and less dusty than the bulk of SDSS galaxies the Millennium cosmological simulation is expected to represent. This is the origin of the differences between the shaded and thick solid histograms in Fig.~\ref{fig:distOR}.

\begin{figure*}
\begin{center}
\includegraphics[width=0.35\textwidth]{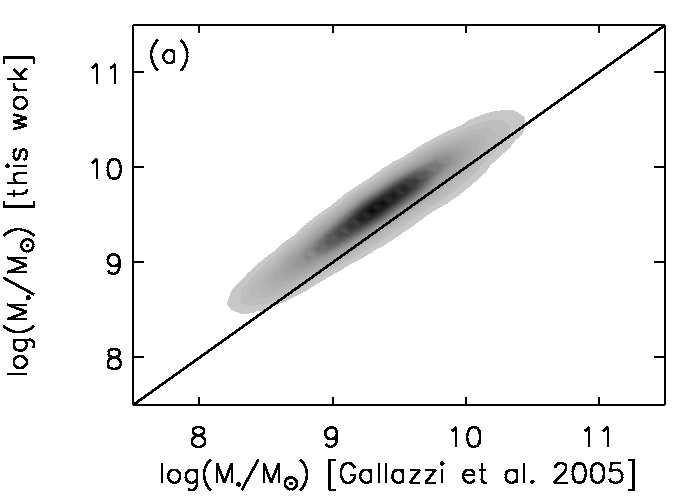}
\includegraphics[width=0.35\textwidth]{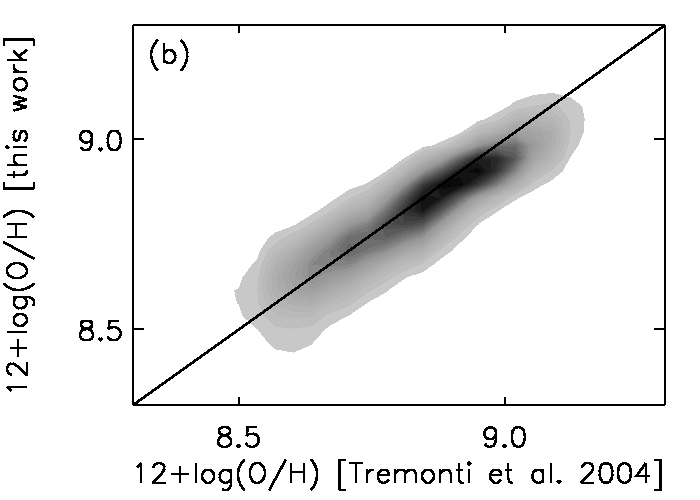}
\includegraphics[width=0.35\textwidth]{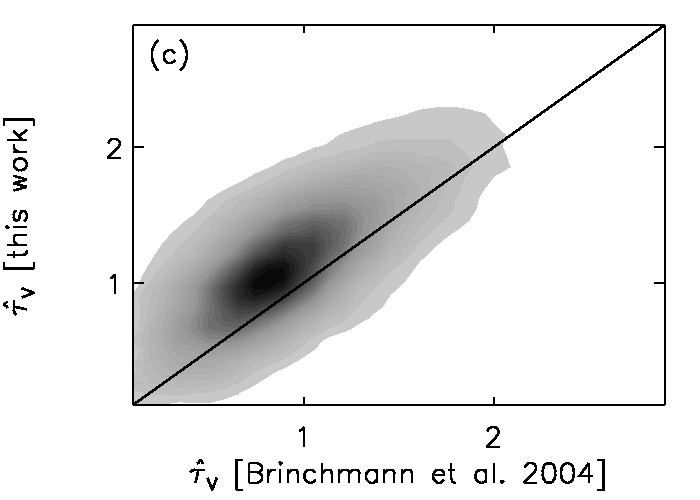}
\includegraphics[width=0.35\textwidth]{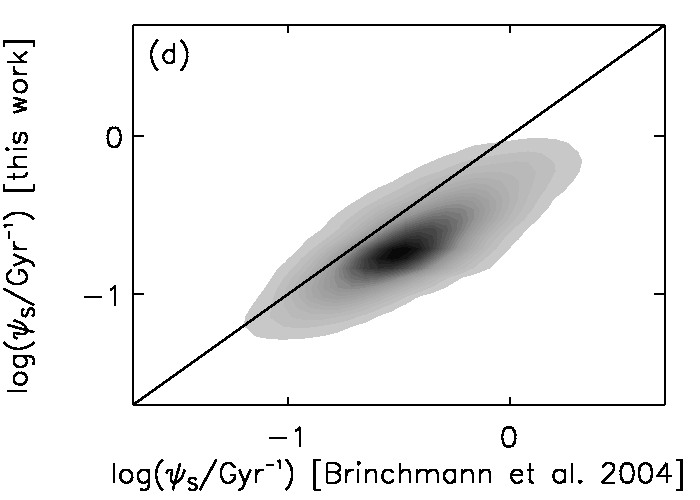}
\caption{Same as Fig.~\ref{fig:r1000jar}, but using the prior distributions in Fig~\ref{fig:distOR} instead of those in Fig.~\ref{fig:dist} to retrieve the probability density functions of \mstar, \aboh, \tauv\ and \ssfr\  for the 12,660 SDSS star-forming galaxies in the sample.}
\label{fig:r1000jarPP}
\end{center}
\end{figure*}

In Fig.~\ref{fig:r1000jarPP}, we show the analog of Fig.~\ref{fig:r1000jar} obtained when using the prior distributions in Fig~\ref{fig:distOR}, instead of those in Fig.~\ref{fig:dist}, to retrieve the probability density functions of \mstar, \aboh, \tauv\ and \ssfr\  for the 12,660 SDSS star-forming galaxies in our sample. The upward bias in the prior distribution of \mlr\ relative to the observed sample (Fig.~\ref{fig:distOR}a) causes the estimates of \mstar\ in Fig.~\ref{fig:r1000jarPP}a to be systematically larger than those derived by \citet{gallazzi2005}. In Fig.~\ref{fig:r1000jarPP}d, the retrieved \ssfr\ is about 60 percent smaller than that estimated by \citet{brinchmann2004}, as in Fig.~\ref{fig:r1000jar}d. The offset in this case results from the downward bias in the prior distribution of \ssfr\ (Fig.~\ref{fig:distOR}c) rather than from differences in the dust attenuation curve, which is assumed to be the same by \citet{delucia2007} and \citet{brinchmann2004}. Meanwhile, the upward bias in the prior distribution of \tauv\ (Fig.~\ref{fig:distOR}e) causes estimates of this parameter in Fig.~\ref{fig:r1000jarPP}c to be systematically larger than those derived by \citet{brinchmann2004}. In Fig.~\ref{fig:r1000jarPP}b, the estimates of \aboh\ retrieved using the prior distribution of Fig.~\ref{fig:distOR}d agree well with those derived by \citet{tremonti2004}.

The results of Figs~\ref{fig:r1000jar} and \ref{fig:r1000jarPP} illustrate the importance of the choice of appropriate prior distributions when applying the Bayesian approach described in Section~\ref{sec:stat} to retrieve physical parameters from observed spectral energy distributions of galaxies. While the prior distributions of Fig.~\ref{fig:distOR} are expected to be representative of the SDSS sample as a whole, adopting these distributions to analyze subsamples of galaxies with properties different from the bulk can induce unwanted offsets in the retrieved physical parameters. In cases where some specific physical property of the observed galaxies can be known in advance (e.g., actively star-forming, early-type, metal-poor), the use of appropriate restrictive prior distributions can help reduce the uncertainties in the retrieved parameters. When no advance information is available, it is preferable to appeal to flat (maximum-ignorance) prior distributions encompassing wide ranges of galaxy physical properties.

\section{Summary and conclusion}
\label{sec:conclusion}

\begin{table*}
\caption{Summary of the constraints on galaxy physical parameters retrieved from different types of optical observations using the method developed in Section~\ref{sec:models}. For each parameter and each type of observation, the quoted uncertainty (defined as half the 16th--84th percentile range of the retrieved likelihood distribution) and accuracy (defined as the absolute difference between the retrieved best estimate and true parameter value) are median quantities determined from the analysis of 10,000 pseudo-observations in Section~\ref{sec:estim}.}
\centering
\begin{minipage}{\textwidth}
\begin{tabular*}{\textwidth}{@{}@{\extracolsep{\fill}}l l c c c c c c c @{}}

\hline
Parameter			& Constraint&Photometry$^a$&\multicolumn{2}{c}{Net emission-line EWs$^b$}&\multicolumn{4}{c}{Spectroscopy$^b$}	\tn
								&	& rest-frame $ugriz$  & $R=100$ & $R=1000$ & \multicolumn{2}{c}{$R=100$} & \multicolumn{2}{c}{$R=1000$}	\tn

								&	&	All$^c$	& 	SF$^d$	& SF$^e$	&All$^c$	&SF$^f$	&All$^c$	&SF$^g$\tn
\hline
$\log[(\mlr)/(\mlrsun)]$\dotfill			&Uncertainty				&0.19		&\dots		&\dots		&0.13	&0.14	&0.10	&0.11	\tn
								&Accuracy				&0.04		&\dots		&\dots		&0.01	&0.01	&0.00	&0.00	\tn
\\[-5pt]
\fsfh\dotfill							&Uncertainty				&0.36		&\dots		&\dots		&0.23	&0.23	&0.17	&0.19	\tn
								&Accuracy				&0.17		&\dots		&\dots		&0.05	&0.05	&0.02	&0.03	\tn
\\[-5pt]
$\log(\ssfr/\mathrm{Gyr}^{-1})$\dotfill		&Uncertainty				&0.67		&0.23		&0.18		&0.24	&0.22	&0.12	&0.12	\tn
								&Accuracy				&0.15		&0.12		&0.05		&0.01	&0.00	&0.00	&0.00	\tn
\\[-5pt]
\aboh\dotfill						&Uncertainty				&0.50		&0.23		&0.05		&0.28	&0.21	&0.13	&0.08	\tn
								&Accuracy				&0.33		&0.11		&0.03		&0.04	&0.03	&0.00	&0.00	\tn
\\[-5pt]
$\tauv$\dotfill						&Uncertainty				&0.91		&\,\,\,0.48$^h$	&\,\,\,0.41$^h$	&0.64	&0.52	&0.35	&0.30	\tn
								&Accuracy				&0.21		&\,\,\,0.57$^h$	&\,\,\,0.14$^h$	&0.05	&0.00	&0.00	&0.00	\tn
\\[-5pt]
$\mu$\dotfill						&Uncertainty				&0.20		&\dots		&\dots		&0.16	&0.15	&0.10	&0.10	\tn
								&Accuracy				&0.10		&\dots		&\dots		&0.04	&0.02	&0.00	&0.00	\tn		
\hline

\end{tabular*}
$^a$ For fixed signal-to-noise ratio $\sn=30$ in all bands.\\
$^b$ For median signal-to-noise ratio per pixel $\snmed=20$ across the wavelength range $\lambda$=3600--7400\,\AA.\\
$^c$ Averaged over all galaxy types.\\
$^d$ For greater-than-3$\sigma$ detections of $\oii\lambda3727$; \hb; $\oiii\lambda\lambda4959,5007$; $\nii\lambda6548+\ha+\nii\lambda6584$; and $\sii\lambda\lambda6716,6731$.\\
$^e$ For greater-than-3$\sigma$ detections of $\oii\lambda3727$; \hb; $\oiii\lambda4959$; $\oiii\lambda5007$; $\nii\lambda6548$; \ha; $\nii\lambda6584$; $\sii\lambda6716$ and $\sii\lambda6731$.\\
$^f$ For galaxies with net \ha+\nii\ emission equivalent width greater than 5\,\AA.\\
$^g$ For galaxies with net \ha\ emission equivalent width greater than 5\,\AA.\\
$^h$ Constraint on the birth-cloud component of the attenuation only (i.e. \tauvbc\ and not \tauv).\\
\end{minipage}
\label{tab:summary}
\end{table*}

We have developed a new approach to constrain galaxy physical parameters from the combined interpretation of stellar and nebular emission in wide ranges of photometric and spectroscopic observations. Our approach relies on the simulation of a comprehensive library of 5 million galaxy spectral energy distributions using a set of state-of-the-art models of star formation and chemical enrichment histories, stellar population synthesis, nebular emission and attenuation by dust. This library can be used to retrieve probability density functions of physical parameters from the Bayesian analysis of any type of multi-wavelength galaxy observation. We focus in this paper on the retrievability of the observer-frame absolute $r$-band stellar mass-to-light ratio (\mlr), the fraction of a current galaxy stellar mass formed during the last 2.5\,Gyr (\fsfh), the specific star formation rate (\ssfr), the gas-phase oxygen abundance [\aboh], the total effective $V$-band absorption optical depth of the dust (\tauv) and the fraction of this arising from dust in the ambient ISM ($\mu$) from different types of observations at optical wavelengths: 5-band $ugriz$ photometry; equivalent-width measurements of strong emission lines; low-resolution spectroscopy (50\,\AA\  FWHM over the wavelength range $\lambda$=3600--7400\,\AA); and medium-resolution spectroscopy (5\,\AA\  FWHM over the same wavelength range).

Since we cannot know the true properties of any sample of observed galaxies, assessing the retrievability of galaxy physical parameters from these different types of observations requires the simulation of pseudo-observations. We therefore simulate pseudo-observations by convolving the spectral energy distributions of models with known parameters with appropriate instrument responses and then applying artificial noise to mimic true observations. For each type of photometric or spectroscopic observation, we simulate 10,000 pseudo-observations in this way and use the other  $\sim5$ million models in the library to retrieve likelihood distributions of \mlr, \fsfh, \ssfr, \aboh, \tauv\ and $\mu$ for each pseudo-galaxy (using equations~\ref{eq:prob}--\ref{eq:weight}). Table~\ref{tab:summary} summarizes the main conclusions of our study. For each parameter and each type of observation, we list the median retrieved uncertainty (defined as half the 16th--84th percentile range of the retrieved likelihood distribution) and accuracy (defined as the absolute difference between the retrieved best estimate and true parameter value) determined from the analysis of the 10,000 pseudo-observations in Section~\ref{sec:estim}.

Several interesting conclusions can be drawn from Table~\ref{tab:summary}. We find that, for example, rest-frame $ugriz$ photometry can provide roughly 50-percent uncertainty on \mlr\ measurements, i.e. almost as tight as the 25- to 35-percent uncertainty provided by medium- and low-resolution optical spectroscopy (photometric constraints worsen somewhat if the galaxy redshift is not known; see Fig.~\ref{fig:photzzz}). Meanwhile, optical spectroscopy is required to constrain \fsfh\ through the strong absorption-line signatures of intermediate-age stars. Table~\ref{tab:summary} also shows that observations of the net emission equivalent widths of prominent optical lines are sufficient to constrain \ssfr\ to within 50--70 percent (depending on the spectral resolution) and \aboh\ to within 0.05--0.23 in star-forming galaxies. Optical spectroscopy can help reduce these uncertainties and the potential small offsets in the retrieved best estimates of \ssfr\ and \aboh. In addition, spectroscopy is required to constrain both parameters in quiescent star-forming galaxies with weak emission lines ($\ssfr\la0.07\,\textrm{Gyr}^{-1}$; see Fig.~\ref{fig:ew}). Medium-resolution spectroscopy provides the best constraints on dust parameters in Table~\ref{tab:summary}, with uncertainties of typically 0.35 in \tauv\ and 0.10 in $\mu$ (these include uncertainties in the optical properties and spatial distribution of the dust as well as in the orientation of the galaxy; see Section~\ref{sec:dust}). At lower spectral resolution, the less distinct signatures of emission lines make the uncertainty in \tauv\ rise by about 80 percent, while the equivalent widths of optical emission lines at any spectral resolution constrain only the part of the attenuation arising from stellar birth clouds.

We stress that the constraints in Table~\ref{tab:summary} rely critically on the ability with our approach to interpret simultaneously the stellar and nebular emission from galaxies. In the case of low-resolution spectroscopy, no reliable constraint can be obtained on any of the physical parameters in Table~\ref{tab:summary} (except for the mass-to-light ratio of early-type galaxies) when nebular emission is not included in the analysis  (Fig.~\ref{fig:specr100deg}). The same is true at medium spectral resolution, although if the resolution is high enough, prominent nebular emission lines can be excised from the spectra and analyzed separately from the stellar continuum and absorption-line signatures. We find that the likelihood distributions of \mstar, \aboh, \tauv\ and \ssfr\ retrieved from the combined analysis of stellar and nebular emission in the spectra of 12,660 SDSS star-forming galaxies (degraded to a resolution of 5\,\AA\  FWHM) using our approach agree well (modulo some recent model improvements) with those obtained in previous separate analyses of the stellar and nebular emission of these galaxies at twice higher spectral resolution \citep[see Fig.~\ref{fig:r1000jar}]{brinchmann2004,tremonti2004,gallazzi2005}. Accounting for nebular emission is less crucial when deriving constraints on \mlr\ from $ugriz$ photometry (Fig.~\ref{fig:photdeg}, bottom row).

Another important factor we can investigate in a straightforward way using our approach is the influence of signal-to-noise ratio on the retrieval of galaxy physical parameters from different types of observations. For multiband photometry, we find that lowering the signal-to-noise ratio from $\sn=30$ to 10 makes the uncertainty in the retrieved \mlr\  in Table~\ref{tab:summary} larger by about 15 percent (Fig.~\ref{fig:photdeg}). For spectroscopy, adopting a median signal-to-noise ratio per pixel $\snmed=5$ instead of 20 makes the median uncertainties in the different retrieved parameters increase by about 30 to 90 percent and introduces substantial biases in the retrieved likelihood distributions (Fig.~\ref{fig:specr100deg}). The ability to make such assessments is valuable as it can help decide between different strategies when designing an observational campaign. For example, as noted in Section~\ref{sec:lowres}, low-resolution spectroscopy with $\snmed=5$ over the wavelength range from $\lambda=3600$ to 7400\,\AA\ provides hardly better constraints on the different parameters in Table~\ref{tab:summary} than high-quality $ugriz$ photometry (Figs~\ref{fig:photbase} and \ref{fig:specr100deg}). In contrast, the constraints from low-resolution spectroscopy with $\snmed=20$ over the same wavelength range are generally far superior to those obtained from multiband photometry (Table~\ref{tab:summary}).

It is important to keep in mind that the results reported in Table~\ref{tab:summary} about the retrievability of galaxy physical parameters from different types of observations correspond to the optimistic case in which the models we rely on are good approximations of true galaxies (Section~\ref{sec:stat}). We have ensured this in the best possible way by appealing to state-of-the-art models including the most recent progress in galaxy spectral modeling to generate the library of spectral energy distributions on which our approach is based. With this in mind, the tool we have developed should be useful not only to interpret existing galaxy datasets, but also to design future observations. In principle, such observations can be of any photometric or spectroscopic type (or a combination thereof ) across the wavelength range covered by spectral evolution models. We have focused in this paper on a few examples of photometric and spectroscopic observations at rest-frame optical wavelengths. In future work, we will present applications of our approach to the interpretation of the ultraviolet and infrared emission from galaxies at various redshifts. The tool we have developed is intended to be made available to the general astronomical community. In the meantime, we encourage colleagues interested in the application of our approach to the interpretation of specific galaxy observations to contact us.


\section*{acknowledgements}

We thank Maryam Shirazi for her help in the assembly of the SDSS galaxy sample. We also thank Marijn Franx, Brent Groves and Vivienne Wild for helpful discussions. This work was funded in part by the Marie Curie Initial Training Network ELIXIR of the European Commission under contract PITN-GA-2008-214227.

Funding for the SDSS has been provided by the Alfred P. Sloan Foundation, the Participating Institutions, the National Science Foundation, the US Department of Energy, the National Aeronautics and Space Administration, the Japanese Monbukagakusho, the Max Planck Society, and the Higher Education Funding Council for England. The SDSS Web site is http://www.sdss.org. The SDSS is managed by the Astrophysical Research Consortium for the Participating Institutions. The Participating Institutions are the American Museum of Natural History, the Astrophysical Institute Potsdam, the University of Basel, Cambridge University, Case Western Reserve University, the University of Chicago, Drexel University, Fermilab, the Institute for Advanced Study, the Japan Participation Group, Johns Hopkins University, the Joint Institute for Nuclear Astrophysics, the Kavli Institute for Particle Astrophysics and Cosmology, the Korean Scientist Group, the Chinese Academy of Sciences, Los Alamos National Laboratory, the Max Planck Institute for Astronomy, the Max Planck Institute for Astrophysics, New Mexico State University, Ohio State University, the University of Pittsburgh, the University of Portsmouth, Princeton University, the US Naval Observatory, and the University of Washington.


\def\aj{AJ}
\def\araa{ARA\&A}
\def\apj{ApJ}
\def\apjl{ApJ}
\def\apjs{ApJS}
\def\apss{Ap\&SS}
\def\aap{A\&A}
\def\aapr{A\&A~Rev.}
\def\aaps{A\&AS}
\def\mnras{MNRAS}
\def\pasp{PASP}
\def\pasj{PASJ}
\def\qjras{QJRAS}
\def\nat{Nature}

\def\aplett{Astrophys.~Lett.}
\def\aas{AAS}
\let\astap=\aap
\let\apjlett=\apjl
\let\apjsupp=\apjs
\let\applopt=\ao

\bibliographystyle{mn2e}
\bibliography{bib_pacifici}

\end{document}